\documentclass{sig-alternate-10pt}
\usepackage[stable]{footmisc}
\usepackage{caption}
\usepackage[normalsize, tight]{subfigure}
\usepackage{graphicx}
\usepackage{float}
\usepackage{color}
\usepackage{epsfig}
\usepackage{url}
\usepackage{tikz}
\usetikzlibrary{arrows}
\usepackage{comment}
\usepackage{todonotes}
\usepackage{diagbox} 
\usepackage{multirow} 
\usepackage{relsize} 
\usepackage{tabularx} 
\usepackage{textcomp}
\usepackage{lmodern}
\usepackage[linesnumbered,ruled,vlined]{algorithm2e}
\usepackage[]{hyperref} 
\definecolor{deepgreen}{RGB}{0,80,0}
\definecolor{darkgreen}{RGB}{0,100,0}
\definecolor{darkred}{RGB}{139,0,0}
\definecolor{darkcyan}{RGB}{0,139,139}
\definecolor{deeppurple}{RGB}{85,26,138}
\hypersetup{
	colorlinks=true,
	citecolor=blue,
	linkcolor=blue
}
\newcommand{\figurename}{Figure~}
\newcommand{\tablename}{Table~}
\newcommand{\figwidththree}{0.35\linewidth}

\newcommand{\rbracket}{]}

\begin{document}
	\title{Micro-burst in Data Centers: Observations, Implications, and Applications}
	\author{
		Danfeng Shan*, Fengyuan Ren*, Peng Cheng*$^\dagger$, Ran Shu* \\
		* Tsinghua University ~~~~~ $^\dagger$ Microsoft Research Asia
	}
	\maketitle
\begin{abstract}
	Micro-burst traffic is not uncommon in data centers.
	It can cause packet dropping, which results in serious performance degradation (e.g., Incast problem).
	However, current solutions that attempt to suppress micro-burst traffic
	are extrinsic and ad hoc, since they lack the comprehensive and essential understanding of micro-burst's root cause and dynamic behavior.
	On the other hand, traditional studies focus on traffic burstiness in a single flow,
	while in data centers micro-burst traffic could occur with highly fan-in communication pattern,
	and its dynamic behavior is still unclear.
	\par To this end, in this paper we re-examine the micro-burst traffic in typical data center scenarios.
	We find that evolution of micro-burst is determined by both TCP's self-clocking mechanism and bottleneck link.
	Besides, dynamic behaviors of micro-burst under various scenarios can all be described by the slope of queue length increasing.
	Our observations also implicate that conventional solutions like absorbing and pacing
	are ineffective to mitigate micro-burst traffic.
	Instead, senders need to slow down as soon as possible.
	Inspired by the findings and insights from experimental observations,
	we propose S-ECN policy, which is an ECN marking policy leveraging the slope of queue length increasing.
	Transport protocols utilizing S-ECN policy can suppress the sharp queue length increment by over 50\%, and reduce the $99^{th}$ percentile of query completion time by $\sim$20\%.
\end{abstract}

\section{Introduction}
\par Broadly speaking, micro-burst is highly intense traffic appearing in a relatively short period.
Recently, micro-bursts in modern data centers attract much attention in both academic
and industry community since they cause serious performance problems.
When massive packets swarm into the same switch port in a twinkling of an eye,
the buffer is sharply bulged, and even overwhelmed.
The resulting high queuing delays and jitters impose heavy penalties on financial trading applications in data centers \cite{financial},
and the packet dropping in succession triggers abnormal timeouts, which causes TCP incast throughput collapse \cite{FAST08incast, SIGCOMM09incast, WREN09incast},
and sluggish response time or low quality of results \cite{SIGCOMM10DCTCP, SIGCOMM11D3, SIGCOMM12D2TCP}.
\par Because of a lack of comprehensive and essential understanding on both root cause and dynamic behavior of micro-burst in data centers,
most of existed solutions to suppress micro-burst and eliminate its negative impact are ad hoc and extrinsic,
for example reducing RTOmin \cite{SIGCOMM09incast}, limiting the number of concurrent flows and adding jitter in application layer \cite{Jeff-talk},
absorbing micro-burst in switch buffer \cite{Arista:Whitepaper2, Broadcom-smart-buffer, Extreme:whitepaper, INFOCOM15Burst},
smoothing micro-burst through pacing packets at sources \cite{NSDI12HULL}.
\par Intuitively, micro-burst traffic stems from various elements,
including data generation pattern in applications \cite{IMC03Burst, CoNEXT13NIC},
system call batching and protocol stack processing in OS \cite{CoNEXT13NIC},
coalescing packets in NICs \cite{IntelRSC, CM10Energy}.
Definitely, the end-to-end congestion control mechanism built-in window-based transport protocol, such as TCP and DCTCP, is the source of a considerable amount of traffic bursts.
Historically, lots of work \cite{SIGCOMM91ACK-compression, ICDCS00burst, IMC03Burst, SIGMETRICS05JIANG, PAM05Allman, CCR05Allman, PAM05burst}
examined the nature of both micro-burst and macro-burst \cite{CCR05Allman} induced by transport protocols in the context of traditional Internet,
and mainly focused on the effect of a single long-lived TCP flow on the formation and dynamic evolution of burst traffic. 
\par The features of modern data center networks are obviously different from the counterparts of traditional Internet, such as symmetrical topology and short network radius.
The traffic distribution and pattern are also unique, for example, the short-lived flows are numerous, but a few long-lived flows monopolize the majority of network resources \cite{IMC09DC}.
The new computing paradigms (such as MapReduce \cite{OSDI08MapReduce} and distributed streaming computing \cite{ICDMW10S4})
and application systems (such as in-memory computing \cite{NSDI13Memcache}, and distributed machine learning \cite{HotCloud15Machine-learning}) require various communication patterns,
which intensify high fan-in bursty traffic \cite{Arista:Whitepaper2}.
Moreover, diverse workflow patterns (such as Partition/Aggregate and Dependent/Sequential \cite{SIGCOMM12DeTail, SoCC12Chronos}) further complicate traffic patterns in data centers. 
Due to the above factors, the conclusions, which describe the nature of micro-burst and macro-burst induced by a single long TCP flow in Internet,
cannot be extended to predicting the dynamic behavior of micro-burst caused by aggregated TCP short flows in data centers,
and thus hardly inspire some reasonable guidelines towards removing its negative effects radically.  
\par Considering the features of traffic pattern and distribution in data centers,
in this work, we re-examine the micro-burst traffic under the control of window-based transport protocol. The main contributions are summarized in three aspects:  
\par (1) Monitoring the evolution of queue length at fine-grained timescales in typical traffic scenarios
and conducting analysis on observations,
we infer the temporal and spatial dynamic behaviors of micro-burst,
and obtain some interesting findings:
1) The self-clocking mechanism and the bottleneck link jointly dominate the evolution of micro-burst;
2) In any traffic scenarios, regardless of with or without long-lived background flows, and including synchronous or asynchronous fan-in short-lived flows,
there is an immutable variable (i.e., the slope of queue length increasing) that can precisely describe the dynamic behaviors of micro-burst in most situations. 
\par (2) Our findings indicate some implications about guidelines towards suppressing micro-burst.
The conventional mechanisms, such as absorbing and pacing, are ineffective in the context of data centers.
Radically, as for latency-sensitive short message transmission in data centers,
the ideal solution is to detect congestion in time and to properly throttle sending rate as soon as possible.
\par (3) Enlightened by our findings and implications,
and leveraging the slope of queue length increasing as an indicator of detecting congestion,
we propose an ECN probability marking policy, called S-ECN,
to eliminate the negative impact of micro-burst caused by aggregated concurrent flows.
Since the slope of queue length can predict the dynamic behavior of micro-burst,
when the S-ECN policy is employed, the queue length increment is effectively suppressed by over 50\% compared with mainstream transport protocols in data centers.
Overall, $99^{th}$ percentile of query completion time can be reduced by $\sim$20\%.
\par The rest of paper is organized as follows.
In \textsection \ref{section:methodology}, we briefly describe the background and present our methodology.
In \textsection \ref{section:characterizing}, we study the micro-burst traffic under different scenarios, and show our experimental observations and findings.
In \textsection \ref{section:solution}, we propose S-ECN policy and evaluate its performance.
Finally, we discuss some related work in \textsection \ref{section:related-work}, and conclude in \textsection \ref{section:conclusion}.
\section{Methodology} \label{section:methodology}
In this section, we discuss the scenarios we need to consider,
and describe how we observe micro-burst traffic.
\begin{figure}[!t]
	\centering
	\includegraphics[width=0.9\linewidth]{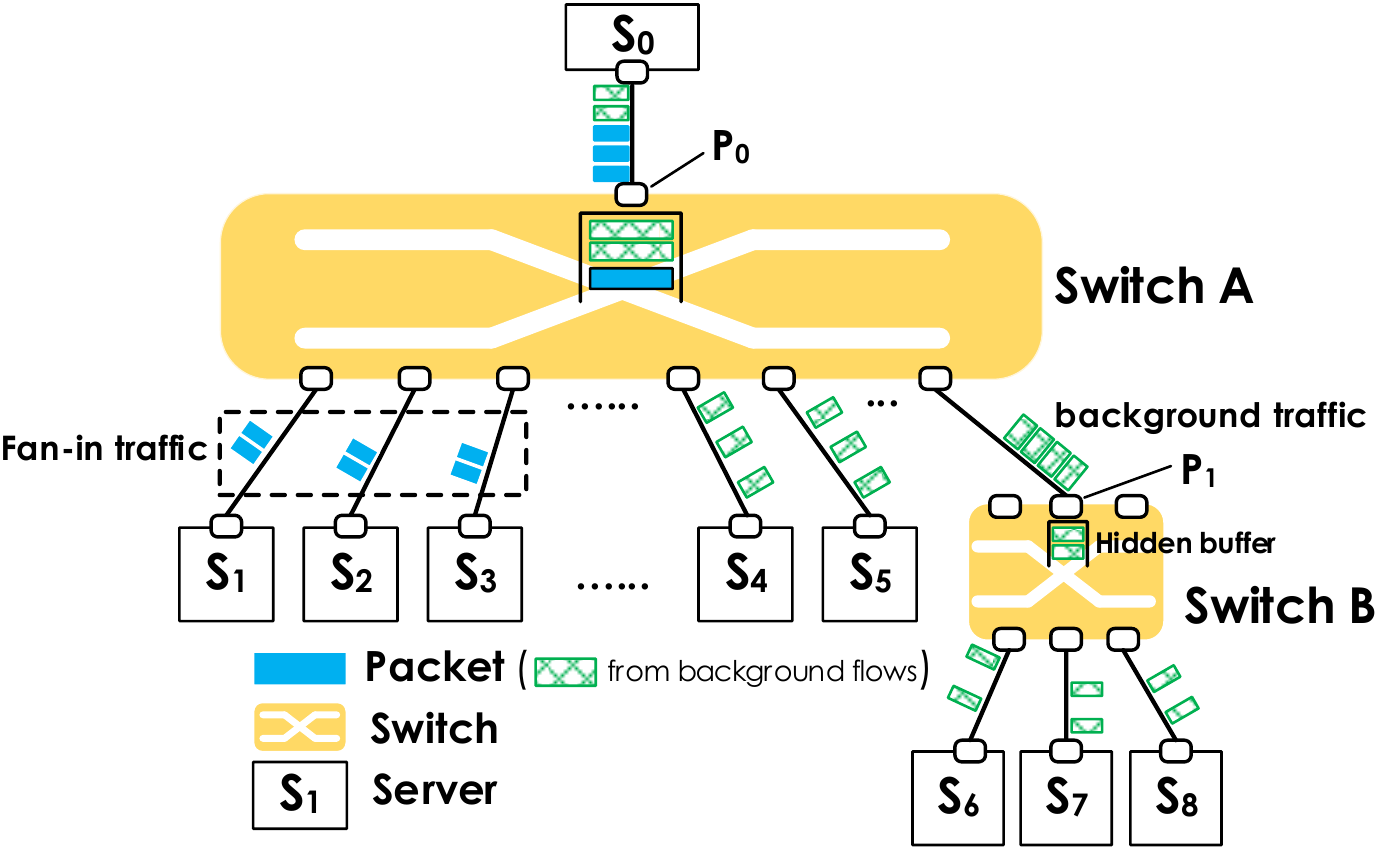}
	\caption{Micro-burst in data centers}
	\label{fig:micro-burst}
\end{figure}

\subsection{Scenario Classification} \label{section:method:scenario}
\par As depicted in \figurename{\ref{fig:micro-burst}},
to comprehensively study the nature and evolution mechanism of micro-burst traffic under the control of transport protocol in data centers,
five typical scenarios are summarized as follows.
\par (1) \emph{Synchronous fan-in traffic:}
In this scenario, many concurrent flows start simultaneously and destine for the same target, as depicted in \figurename{\ref{fig:micro-burst}}.
This scenario is common in many data center application systems.
For example, in distributed storage system \cite{FAST08incast}, a client will synchronously read data blocks from lots of storage servers.
In user-facing online services employing partition-aggregate structure \cite{SIGCOMM10DCTCP}, 
a parent server will query results from many leaf servers in parallel,
and leaf servers will synchronously sending responses to the parent server.
In memcached systems \cite{NSDI13Memcache}, a web server will communicate with many memcached servers to satisfy a user request.
\par (2) \emph{Asynchronous fan-in traffic:}
Different from previous scenario, in this scenario, many concurrent flows start asynchronously, but destine for the same target.
This scenario also exists in multiple data center applications.
For example, in distributed machine learning systems \cite{HotCloud15Machine-learning}, lots of servers frequently communicate with a centralized server,
and thus communications from different sources easily overlap although they are asynchronous.
And in MapReduce framework \cite{OSDI08MapReduce}, many mappers will transfer intermediate results to reducers,
and these transmissions may be asynchronous but can still be bottlenecked together.
\par (3) \emph{Fan-in traffic with one background flow:}
In this scenario, there is one long-lived background flow before fan-in flows start.
The background flow will share the same bottleneck with fan-in flows (congested at port $P_0$ in \figurename{\ref{fig:micro-burst}}).
\par (4) \emph{Fan-in traffic with several background flows congested at the same hop:}
As depicted in \figurename{\ref{fig:micro-burst}}, in this scenario, several background flows have already congested at the port $P_0$ before fan-in flows start.
Different from previous scenario, when fan-in flows start, they can see buffer occupancy caused by background flows at port $P_0$,
which may affect the dynamic behavior of micro-burst traffic.
\par (5) \emph{Fan-in traffic with several background flows congested at previous hop:}
As depicted in \figurename{\ref{fig:micro-burst}}, in this scenario,
several background flows are congested at port $P_1$ in the beginning.
When fan-in flows arrive at port $P_0$, they cannot see any buffer occupancy.
However, as the congestion point shifts from port $P_1$ in Switch B to port $P_0$ in Switch A,
the buffer occupancy in the port $P_1$ will move to the port $P_0$ in a very short time,
which may intensify the traffic burstiness at port $P_0$.
\subsection{Observing Micro-burst}
\begin{figure}[!t]
	\centering
	\subfigure[Synchronous arrival]{
		\includegraphics[width=0.45\linewidth]{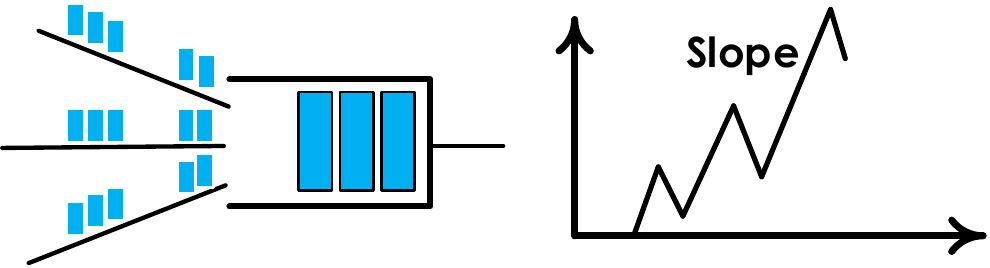}
		\label{fig:method:qevo-mb-syn}
	}
	\hfil
	\subfigure[Asynchronous arrival]{
		\includegraphics[width=0.45\linewidth]{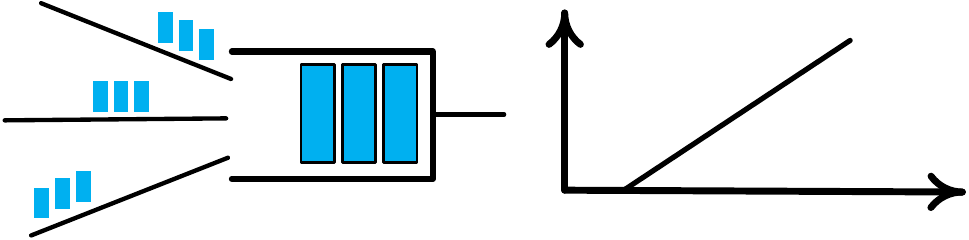}
		\label{fig:method:qevo-mb-even}
	}
	\caption{Characteristics of micro-burst traffic can be inferred from queue length evolution.}
	\label{fig:method:qevo-mb}
\end{figure}
\par Capturing or monitoring micro-bursts in data centers is challenging \cite{cisco-micro-burst-monitor, NSDI11MB},
since switches operate at very high speed (10/40Gbps) while micro-burst occurs in very short timescales (hundreds of microseconds).
On the other hand, through separately studying individual micro-burst on each link or at end systems,
we cannot directly gain the dynamic behavior of micro-burst when lots of flows aggregate,
which is significant since micro-burst traffic causes problems mainly when fan-in traffic causes fast increasing of queue length.
\par In this paper, we study micro-burst traffic through observing evolution of queue length in switch,
which directly reflects the aggregation behavior of traffic.
Furthermore, since queue length in switch evolves differently under different scenarios,
the dynamic behavior of micro-burst caused by aggregation of flows can be inferred from queue length evolution,
as shown in \figurename{\ref{fig:method:qevo-mb}}.
For example, when fan-in flows start synchronously, 
bursts from different flows may arrive simultaneously.
As a result, there are impulses during queue length increasing (\figurename{\ref{fig:method:qevo-mb-syn}}).
When fan-in flows start asynchronously,
bursts from different flows may arrive separately.
As a result, queue length increases in a smoother way
(\figurename{\ref{fig:method:qevo-mb-even}}).
Besides, queue length evolution is also different when there are background flows.
Moreover, as depicted in \figurename{\ref{fig:method:qevo-mb-syn}},
slope of queue increasing (i.e., queue length increasing rate) can reflect the fan-in degree,
since with more concurrent flows the queue length tends to increase faster.
\subsection{Monitoring Queue Length}
\begin{table}[!t]
	\centering
	\begin{tabular}{|c|c|c|c|c|} \hline
		\bf \multirow{2}{*}{Resources}
		& \bf Ref. 				& \bf ECN 					& \bf \multirow{2}{*}{+TPP} 	& \bf \multirow{2}{*}{+S-ECN} \\
		& \bf {\footnotesize Switch}	& \bf {\footnotesize Switch} & & \\ \hline
		{\scriptsize Slice Flip Flops}	
		& 14158					& 14378						& 	14777				& 14700 \\ \hline
		LUTs	& 17589					& 18048						&	19050				& 18544 \\ \hline
	\end{tabular}
	\normalsize
	\caption{Resource usage in NetFPGA. ECN switch is built upon reference switch.
		TPP switch and S-ECN switch are built upon ECN switch.}
	\label{tab:netfpga}
\end{table}

\par To observe micro-burst traffic in switch queue,
we need to get queue length at fine-grained timescales,
because the queue length changes every a few microseconds.
Existing techniques in commodity switches can only monitor queue length at a coarse timescale.
Thus, we develop a method similar to TPP \cite{SIGCOMM14TPP},
which can acquire queue length at per-packet granularity,
and introduce little overhead to switch in the meanwhile.
Specifically, whenever a packet is passing through the switch,
we let it carry the timestamp and current queue length in its payload.\footnotemark~
When the packet arrives at the receiver,
the receiver extracts timestamp and queue length from it,
and stores the information into disk.
\footnotetext{In order not to affect normal traffic (such as ssh) that carries useful data in the payload,
	only packets, whose port value (both source port and destination port) in TCP header is larger than 1023, are used to carry queue length.}
\par We implement the approach on NetFPGA platform \cite{NetFPGA}.
Specifically, we use a 16-bit register to record the queue length in each port,
which is at an 8-byte granularity.
We also use a 32-bit register to maintain timestamp,
which is at an 800-ns granularity (i.e., the register increases every 800ns).
Timestamp and queue length are put into packets before they enter the output queue,
which only need additional processing time of 1 cycle.
Since the clock run at 125MHz, this implementation only adds 8ns latency to the pipeline.
Our implementation needs less than 6\% extra resources, as shown in \tablename{\ref{tab:netfpga}}.

\section{Experimental Observations} \label{section:characterizing}
In this section, we separately study micro-burst traffic in each scenario listed in \textsection \ref{section:method:scenario}.
After that, we present our observations and implications.
\subsection{Testbed} \label{section:testbed}
\begin{figure}[!t]
	\centering
	\includegraphics[width=0.9\linewidth]{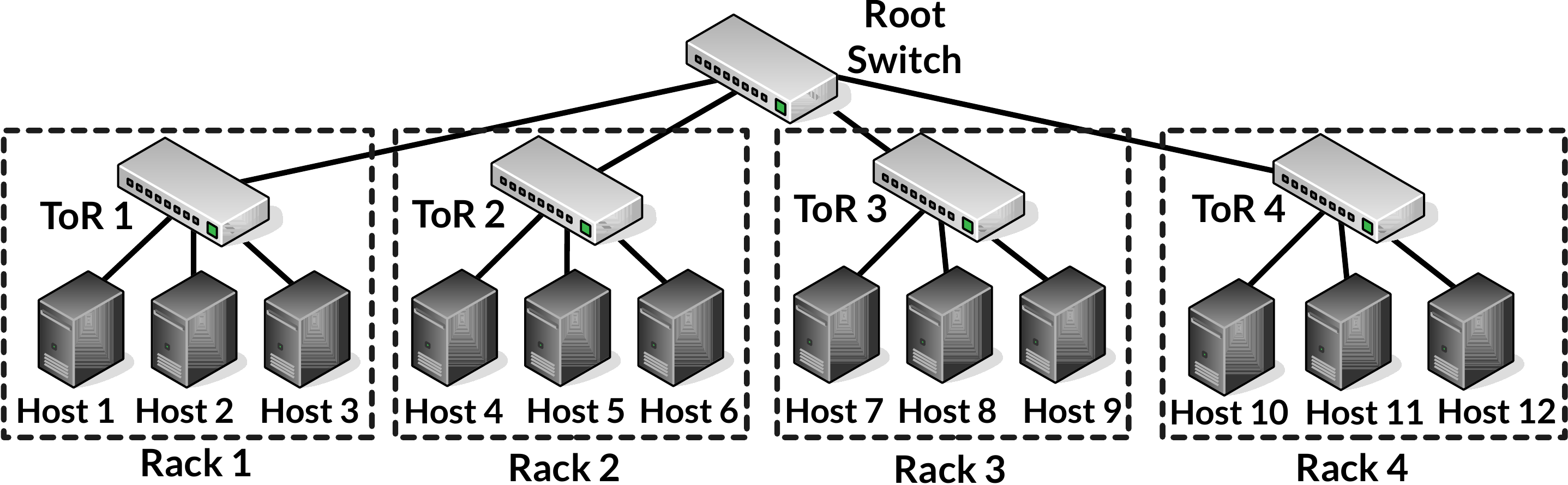}
	\caption{Network topology in experiments}
	\label{fig:topology}
\end{figure}
The topology of our testbed is shown in \figurename{\ref{fig:topology}}.
The testbed contains 12 hosts across 4 racks.
Each rack holds 3 hosts connected to a top-of-rack (ToR) switch.
All ToR switches are connected through a root switch.
Each host is a Dell Optiplex 780 desktops
with an Intel\textregistered~Core\texttrademark~2 Duo E7500 2930 MHz CPU, 4 GB memory,
a 500GB hard disk, and an Intel\textregistered~82567LM Gigabit Ethernet NIC,
running CentOS 5.11 with GNU/Linux kernel 2.6.38.
All switches are NetFPGA cards with four Gigabit Ethernet networking ports.
Buffer size in each output port can be arbitrarily set between 1KB and 512KB.
RTT between hosts from different racks is about 50$\mu$s without queuing.
The queue length in root switch is monitored.
\subsection{Observations and Analysis}
\subsubsection{Synchronous fan-in traffic} \label{section:characterizing:syn}
\begin{figure}[!t]
	\centering
	\subfigure[Queue length evolution]{
		\includegraphics[width=0.9\linewidth]{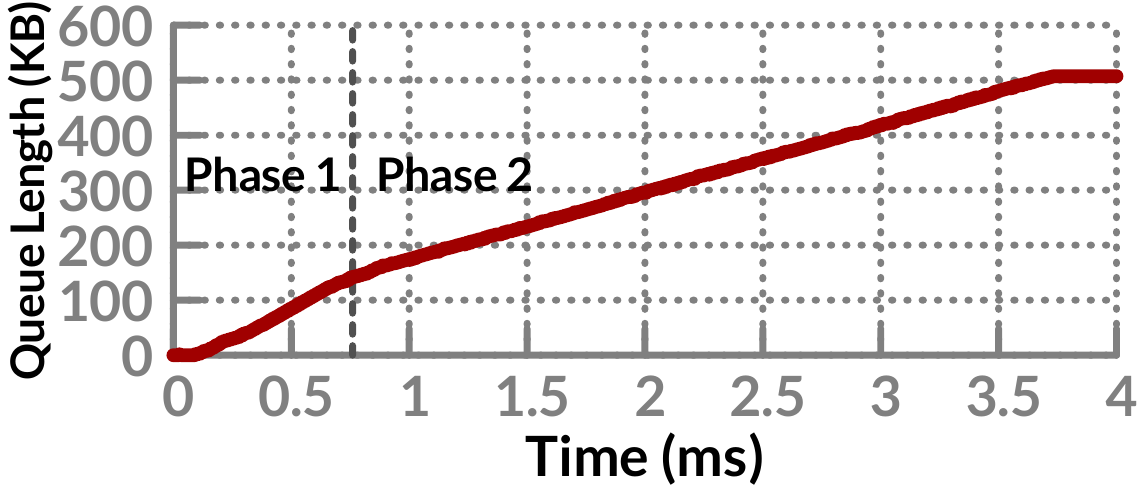}
		\label{fig:syn:qevo}
	}
	\\
	\subfigure[Distribution of slope]{
		\includegraphics[width=0.9\linewidth]{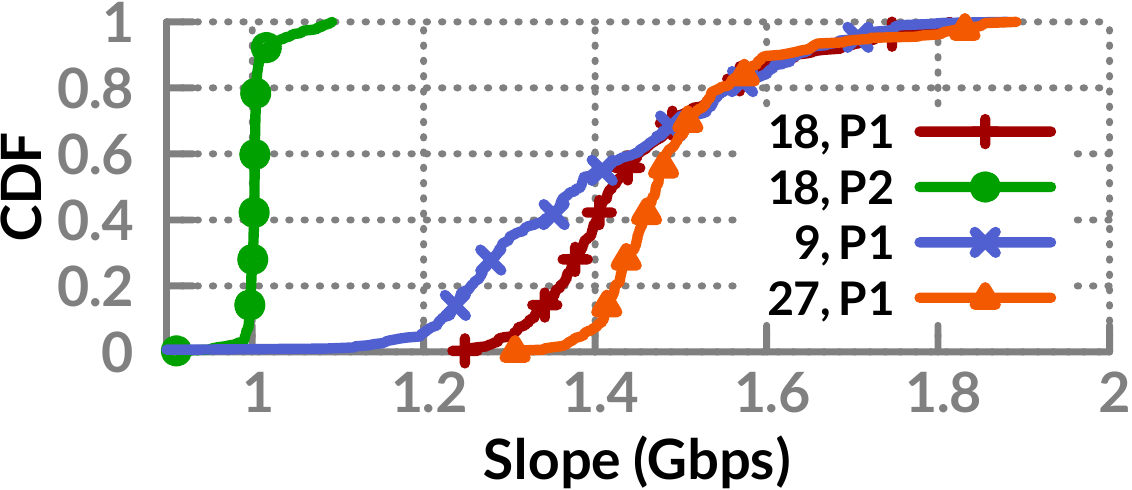}
		\label{fig:syn:slope-cdf}
	}
	\caption{Experiment results in synchronous fan-in scenario. In (b), ``18,P1'' stands for slope in Phase 1 when there are 18 flows.}
\end{figure}
\noindent \textbf{Experiment settings:} Firstly, we let Host 10 send queries to Host 1-9 at the same time
and these hosts will simultaneously send a respond message to Host 10.
Each host is used to emulate multiple senders.
Query size is 50B and response message size is 1000KB.
The buffer size in each output port is set to 512KB.
End hosts use TCP NewReno as their transport protocols\footnotemark.
Large Segment Offload is disabled,
but it will not affect our main results (see discussion in \textsection \ref{section:discussion}).
Delayed ACK is also disabled \cite{NSDI11YU}.
We use default values for other TCP parameters.
\footnotetext{We have studied different TCP variants (Reno, SACK, CUBIC) and observed the similar results.}
\begin{figure}[!t]
	\centering
	\includegraphics[width=0.762\linewidth]{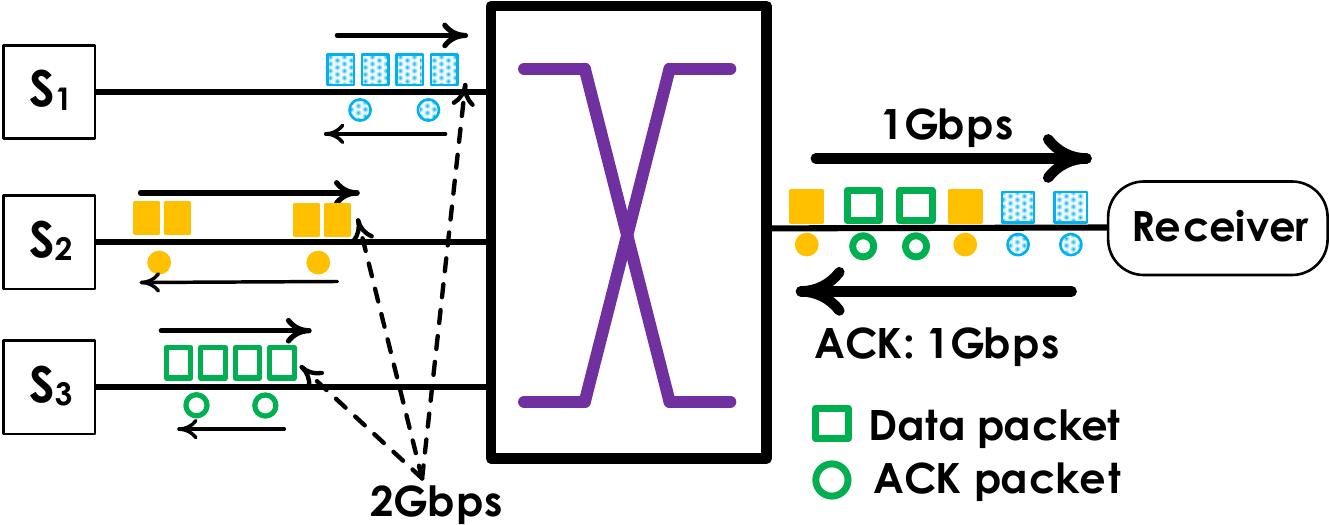}
	\caption{Explanation of Phase 2}
	\label{fig:explain:constant_rate}
\end{figure}
\begin{figure}[!t]
	\centering
	\includegraphics[width=0.8\linewidth]{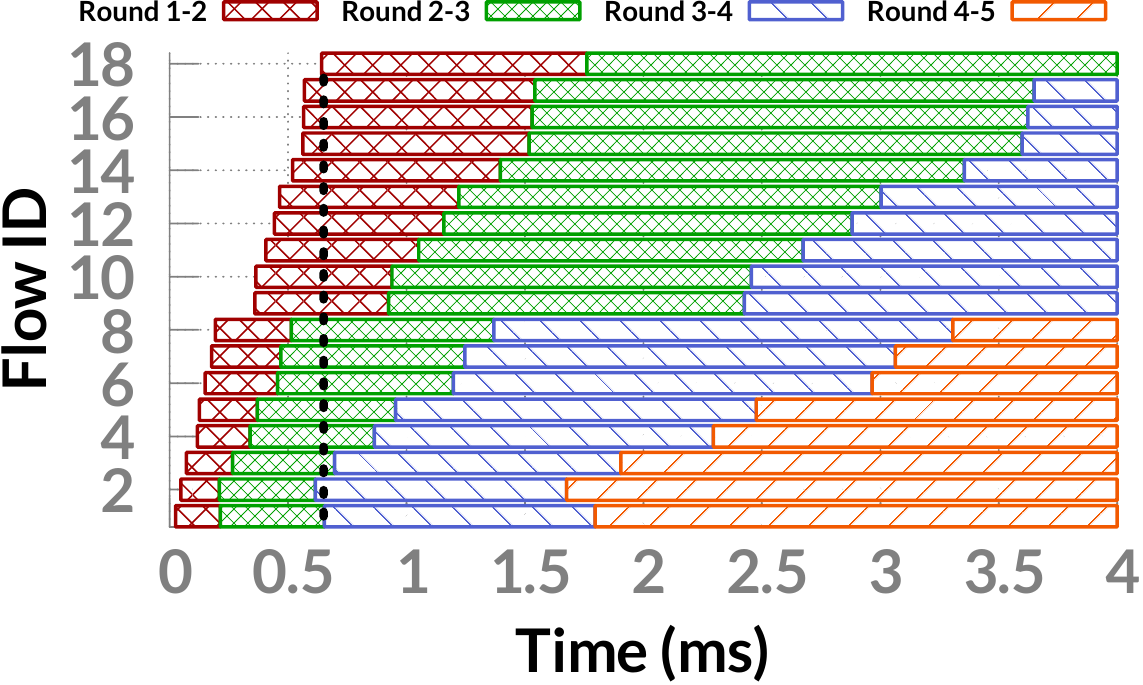}
	\caption{Time interval between two successive rounds of packets}
	\label{fig:explain:round-interval}
\end{figure}
\\\textbf{Experiment results:} \figurename{\ref{fig:syn:qevo}} depicts the queue length evolution in the congested port when there are 18 senders.
The queue length increasing has two phases:
sharply increasing for a relatively short period at the beginning (Phase 1)
and slower increasing after that (Phase 2).
Particularly, in Phase 2, the queue length increases at a constant rate,
without any impulses or jitters.
\par We repeat the experiment for 100 times
and calculate the slope in Phase 1 and Phase 2, respectively.
The slope distributions are shown in \figurename{\ref{fig:syn:slope-cdf}},
where we have another observation:
in Phase 2, \emph{the queue length is not only increasing at a constant rate
	but the increasing rate is always 1Gbps.}
Specifically, when there are 18 flows, the average slope is 1.005Gbps,
and the coefficient of variation is only 0.018.
Through detailed analysis, we find that this phenomenon is caused by
TCP's self-clocking system and evolution of congestion window in slow start phase,
as depicted in \figurename{\ref{fig:explain:constant_rate}}.
Specifically, since data packets are received at 1Gbps,
the receiver is always acknowledging data at 1Gbps,
which is the speed of congested port.
On the other hand,
since all flows are in slow start phase,
each sender introduces two data packets into network when receiving an ACK packet (note that delayed ACK is disabled).
As a result, the total generated traffic from all senders is 2Gbps.
Thus, the overall queue length increasing rate is 1Gbps.
Furthermore, at packet granularity, ACK packets are evenly spread across bottleneck link.
Since the topology is symmetric, after these ACKs arrive at senders, they will evenly trigger transmissions of data packets,
and thus data packets are evenly spread across input links.
As a result, after these packets aggregate, the queue length in congested port is smoothly increasing at constant rate.
\par As shown in \figurename{\ref{fig:syn:slope-cdf}}, in Phase 1, the slope is much larger than that in Phase 2;
and with more concurrent flows, the slope in Phase 1 tends to be larger.
This is because sources are sending the first round of packets in this phase,
and the self-clocking system does not take effect.
To prove this, we find out the ending time of first round among all flows.
\figurename{\ref{fig:explain:round-interval}} shows the time interval between ending time of two successive rounds.
Specifically, the latest ending time of transmitting first round packets is at 0.642ms,
which is in accordance with the ending time of Phase 1 in \figurename{\ref{fig:syn:qevo}}.
\subsubsection{Asynchronous fan-in traffic}
\noindent \textbf{Experiment settings:} In this experiment, 18 flows randomly start during a period of 2ms, which are from Host 1-9 to Host 10.
Other settings keep unchanged.
\begin{figure}[!t]
	\centering
	\subfigure[Queue length evolution]{
		\includegraphics[width=0.9\linewidth]{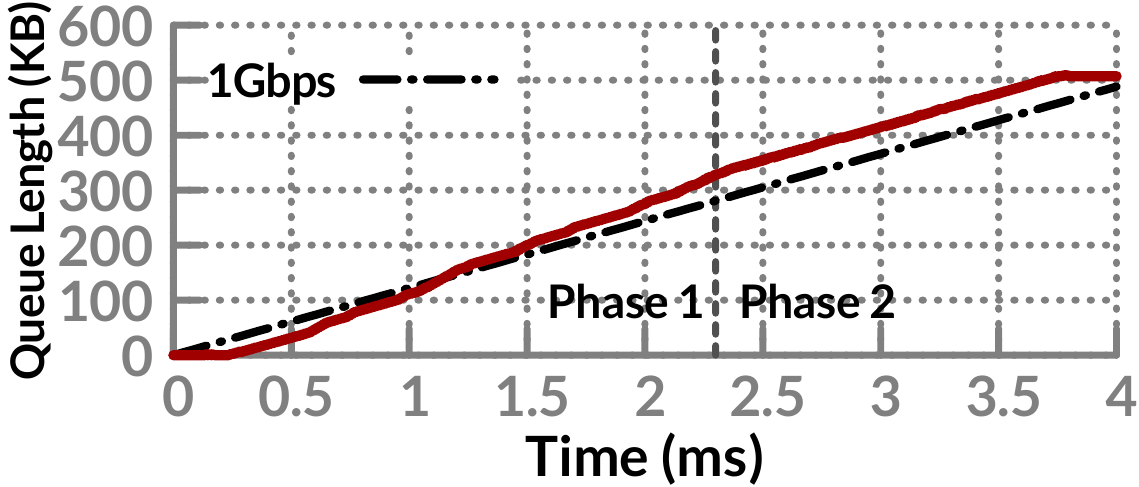}
		\label{fig:asyn:qevo}
	}
	\subfigure[Distribution of slope]{
		\includegraphics[width=0.93\linewidth]{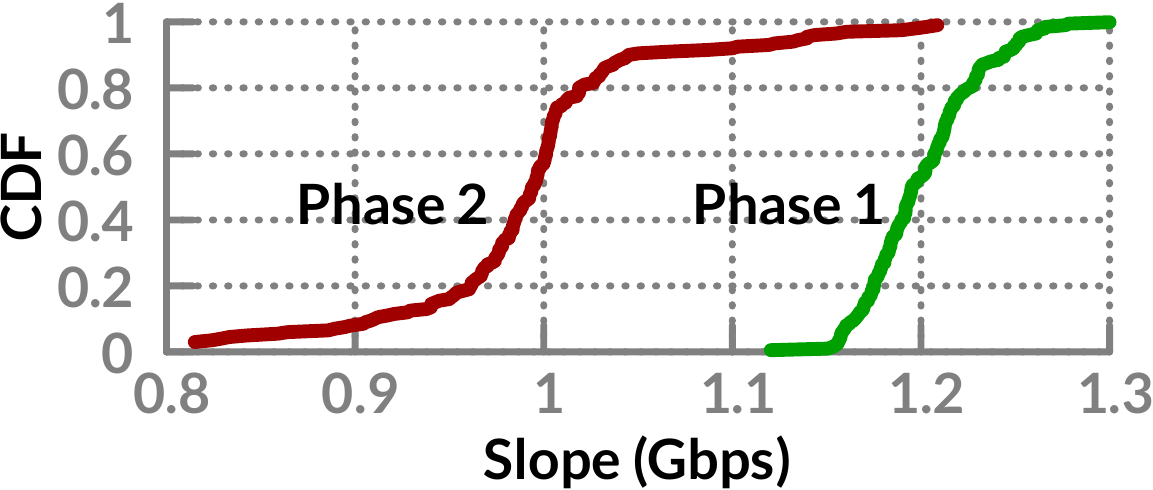}
		\label{fig:asyn:slope-cdf}
	}
	\caption{Experiment results in asynchronous fan-in scenario.}
	\label{fig:asyn}
\end{figure}
\\\textbf{Experiment results:} The queue length evolution is shown in \figurename{\ref{fig:asyn:qevo}}.
Because first rounds of packets from different senders asynchronously arrive at the congested port,
in Phase 1 the queue length does not increase as sharply as that in previous scenario where these packets arrive synchronously.
However, in Phase 2, the queue length is smoothly increasing at a constant rate of port speed,
which is similar to that in synchronous.
This is because all flows have finished sending the first round of packets,
and all packet departures in this phase are triggered by ACKs.
In addition, as in previous scenario, ACKs are evenly acknowledging data at 1Gbps,
and each ACK packet triggers transmissions of two packets since flows are in slow start phase.
Therefore, traffic arriving rate at the congested port is 2Gbps and slope is 1Gbps.
\subsubsection{Fan-in traffic with one background flow}
\noindent \textbf{Experiment settings:} Along with query traffic, we add one background flow.
Specifically, at the beginning,  Host 9 begins to send data to Host 11 as fast as it can.
After 0.5 seconds, Host 10 begins to query Host 1-8.
Other settings are not changed.
\begin{figure}[!t]
	\centering
	\includegraphics[width=0.9\linewidth]{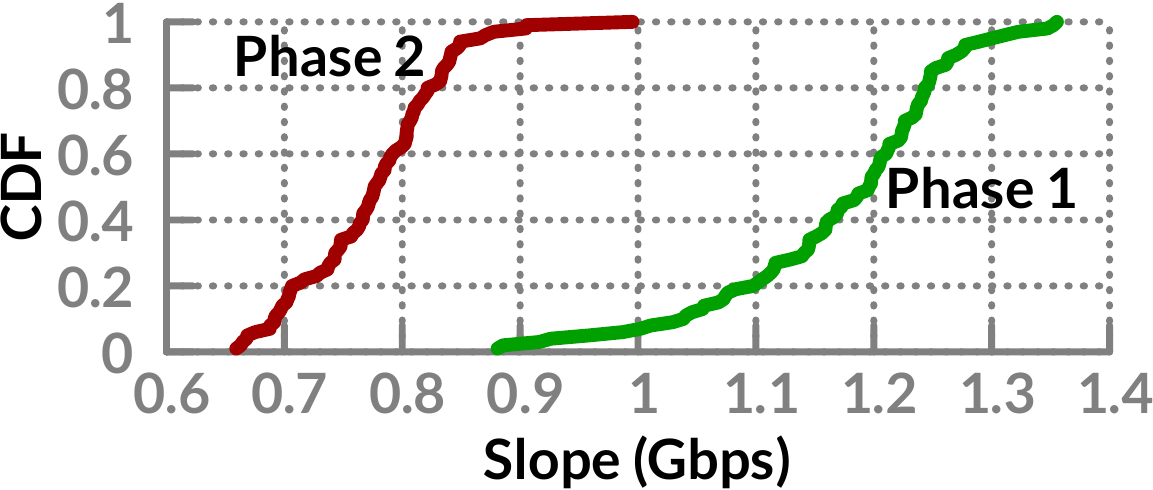}
	\caption{Distribution of slope when there is one background flow.}
	\label{fig:qevo-one-back}
\end{figure}
\\\textbf{Experiment results:} The experiment is repeated for 100 times.
As shown in \figurename{\ref{fig:qevo-one-back}},
the slope in Phase 1 is larger than the port speed as in previous scenarios,
but the slope in Phase 2 becomes much smaller than the port speed.
This is because the congestion window of the background flow has reached its maximum value by the time other fan-in flows start,
and thus the background flow only introduces one data packet into network when receiving an ACK packet\footnotemark.
On the other hand, the other flows inject two data packets when receiving an ACK. 
Overall, the arriving rate of total traffic is less than 2Gbps,
and thus slope is smaller than 1Gbps.
\footnotetext{
	Actually, at the beginning, the sender of the background flow keeps sending data at line rate.
	Because the maximum congestion window (which is usually limited by TCP's send buffer size) can be larger than the pipeline capacity ($C \times RTT$).
	But this has little effect as long as TCP's send buffer is not too large.
	Therefore, we don't discuss it in this paper.
}
\subsubsection{Fan-in traffic with several background flows congested at the same hop} \label{section:characterizing:back}
\begin{figure}[!t]
	\centering
	\subfigure[Queue length evolution.]{
		\includegraphics[width=0.9\linewidth]{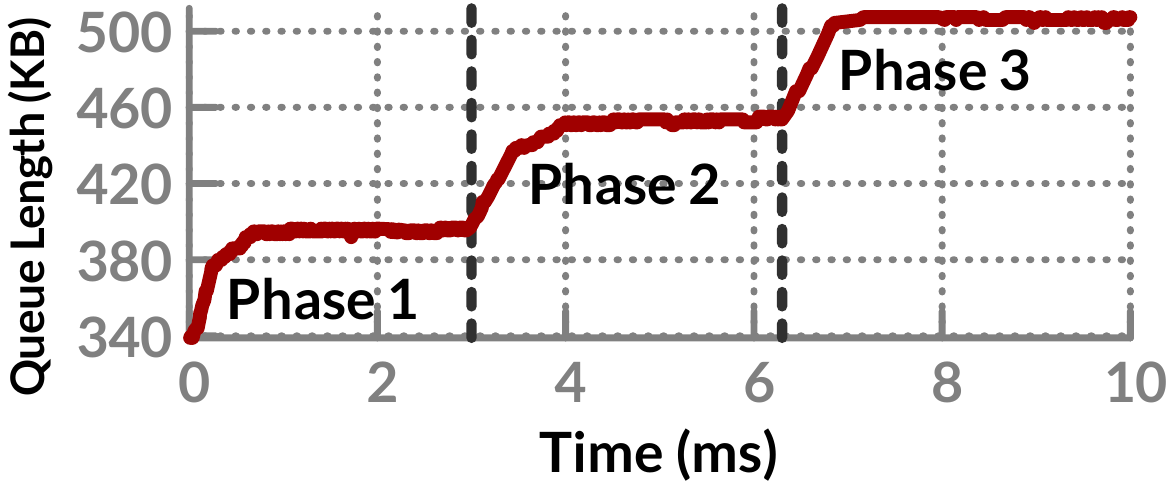}
		\label{fig:qevo-back}
	}
	\centering
	\subfigure[Distribution of slope.
	"P2" stands for "Phase 2".]{
		\includegraphics[width=0.93\linewidth]{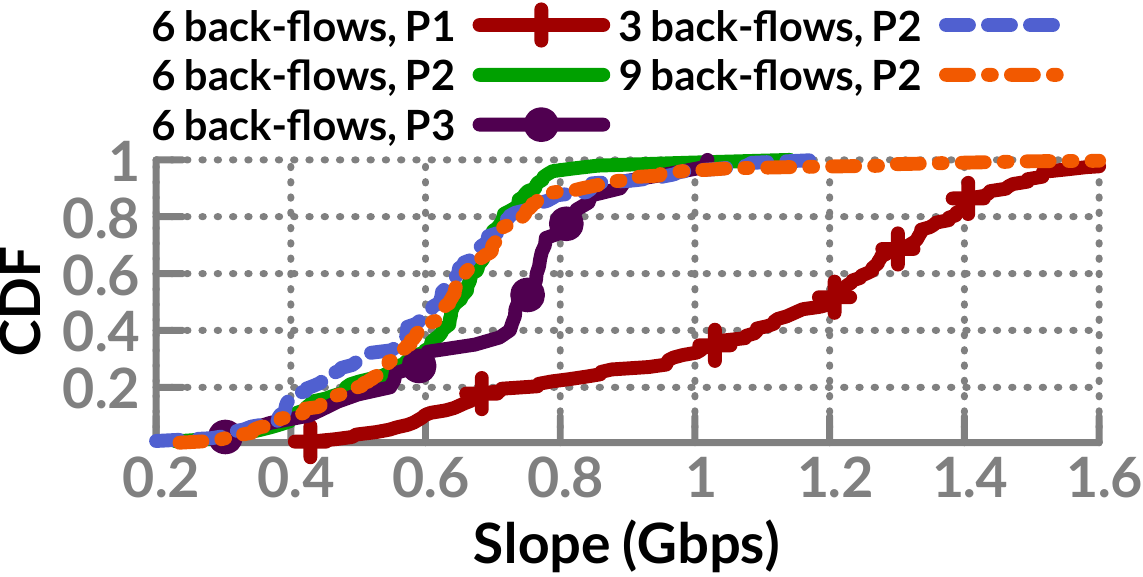}
		\label{fig:w-back:distribution}
	}
	\caption{Experiment results when there are several background flows congested at the same hop.}
\end{figure}

\noindent \textbf{Experiment settings:} Along with query traffic, we add three background flows,
which are from Host 1, 4, 7 to Host 11 and Host 12.
After 0.5 seconds, Host 10 begins to query hosts in other racks (excluding Host 1, 4, 7)
and these hosts send a response message to Host 10.
\\\textbf{Experiment results:} The queue length evolution is shown in \figurename{\ref{fig:qevo-back}}.
The queue length increasing can be split into multiple phases.
Before the fan-in flows start, there is buffer occupancy caused by background flows in the root switch.
In Phase 1, first round of packets from the fan-in flows arrive at the port simultaneously,
therefore the queue length  experiences a sharp increasing.
After that, since the packets from fan-in flows are queued behind those from background flows,
fan-in flows will defer injection of new packets into network,
and the queue length stops sharply increasing.
Similarly, in Phase 2, the queue length increases sharply at the beginning, which is caused by the second round of packets,
and the queue length keeps smoothly increasing after that.
Finally, fast queue length increasing in Phase 3 is caused by the third round of packets.
\par We repeat the experiment for 100 times,
and calculate the slope in each phase when queue length is sharply increasing,
which is shown in \figurename{\ref{fig:w-back:distribution}}.
We have three main observations.
\emph{First,} we find that slope after Phase 1 (i.e., Phase 2 and Phase 3) is much smaller than that without background flows.
This is because some of the arriving packets in these phases are from background flows,
and these background flows are in congestion avoidance phase, in which the congestion window is increased by one MSS every round trip time.
Therefore, although ACKs are acknowledging data at 1Gbps,
data packets are sent at a rate smaller than 2Gbps.
\emph{Second,} after Phase 1, the slope in later phase (e.g., Phase 3) is larger than that in earlier phase  (e.g., Phase 2).
This is because ACKs from slow start flows introduce more packets into network than those from flows in congestion avoidance phase,
and the ratio of packets from slow start flows becomes increasingly large.
As a result, in the reverse path, the ratio of ACKs from slow start flows also becomes increasingly large.
Finally, after these ACKs arrive at sender, the total packet departure rate is larger in later phase.
\emph{Third,} the slope distributions are similar with different number of background flows.
For example, the average slopes are (0.613Gbps, 0.622Gbps, 0.641Gbps) when there are (3, 6, 9) background flows.
This is because the ratio of packets from slow start phase has little change as the number of background flows varies,
and the congestion window increasing rate from background flows can be ignored compared to that from slow start flows.
\subsubsection{Fan-in traffic with several background flows congested at previous hop}
\noindent \textbf{Experiment settings:} In this experiment, the background flows are from Host 7, 8, 9 to Host 11 and Host 12,
thus, these background flows are congested in ToR switch 3.
After 0.5 seconds, Host 10 begins to query Host 1-6 and they send a respond message to Host 10.
At this time, the congestion point will shift to the next hop in root switch.
Other experiment settings keep unchanged.
\begin{figure}[!t]
	\centering
	\includegraphics[width=0.9\linewidth]{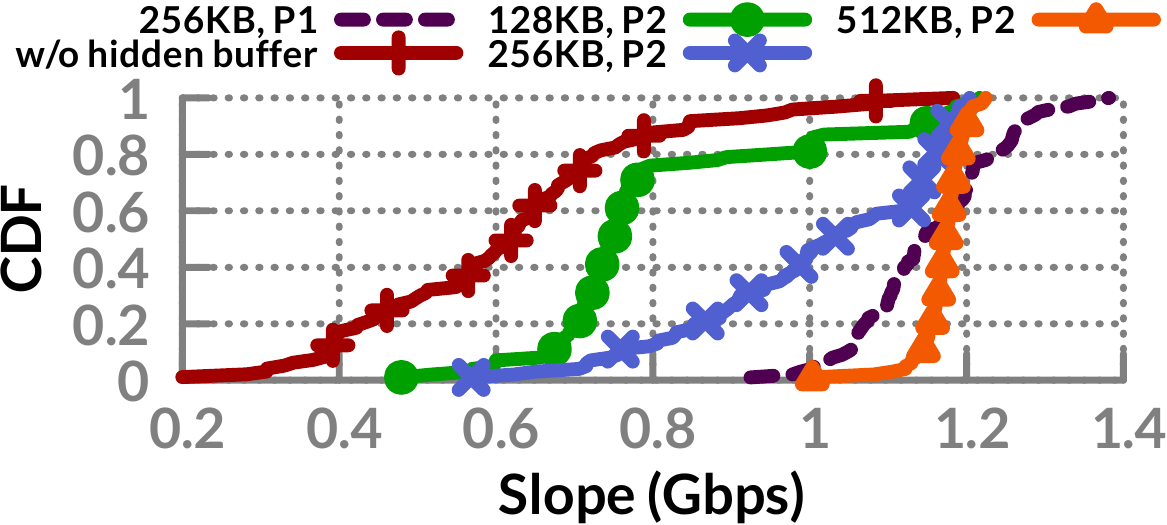}
	\caption{Slope distribution when there is hidden buffer. "P2" stands for "Phase 2".}
	\label{fig:hidden-buffer:distribution}
\end{figure}
\\\textbf{Experiment results:} We repeat each experiment for 100 times.
\figurename{\ref{fig:hidden-buffer:distribution}} shows the distribution of slope.
We compare the slope distribution with that in previous experiment in which the congestion point does not change.
Experiment results show that in this scenario slope in Phase 2 is much larger than that without shift of congestion point,
and may even be larger than the port speed.
This phenomenon is caused by the buffer occupancy in previous hop.
Specifically, in Phase 2,
arriving packets at the root switch are from fan-in flows and background flows.
The packet transmissions from fan-in flows follow self-clocking mechanism.
However, the transmissions of packets from background flows are not determined by self-clocking mechanism,
since they are from buildup queue in the previous hop (i.e., ToR switch 3),
and they are sent at the port speed of ToR switch 3.
As a result, the queue length increasing is intensified by these packets.
Since the buffer occupancy in previous hop cannot be seen by fan-in flows at the beginning,
we call it \emph{hidden buffer}.
\par We examine the effect of hidden buffer by setting the buffer size of ToR switch 3 to 128KB, 256KB, and 512KB, respectively.
As shown in \figurename{\ref{fig:hidden-buffer:distribution}},
the slope increases as the amount of packets in hidden buffer increases.
Specifically, when buffer sizes are 128KB, 256KB, and 512KB, the average slopes are 0.804Gbps, 1.009Gbps, and 1.170Gbps, respectively.
\par In theory, the slope is no larger than  $2aR/(a+R)$,
where $a$ is the arriving rate of first round packets from fan-in flows and $R$ is the port speed.
This is because after 1st round packets arrive at congested port,
the fraction of bandwidth occupied by fan-in flows is about $aR/(a+R)$.
Therefore, in the next round, these fan-in flows will generate traffic at $2aR/(a+R)$ in total.
Meanwhile, sending rate of traffic from hidden buffer to the congested port is at most $R$.
Therefore, the queue length increasing rate is no larger than $2aR/(a+R)$.
\par Note that although in our experiments the packets in hidden buffer only come from one input port,
results can be extended to the scenario in which the packets in hidden buffer come from several input ports,
because total arriving rate of traffic from hidden buffer is no larger than the port speed.
Otherwise, the bottleneck is not in the previous hop.
In particular, the upper bound of slope is also $2aR/(a+R)$.
\subsection{Summary}
\par In summary, when multiple flows aggregate in a queue,
the queue length increasing can be divided into two phases.
In Phase 1, queue length increasing is caused by arriving of 1st round packets.
The slope in Phase 1 directly reflects the communication pattern and intensity.
Specifically, when communications are synchronous, or more concurrent flows aggregate in a queue,
the slope in this phase should be larger.
\par In Phase 2, we have two main observations.
From a macroscopic perspective, the arriving traffic rate is limited by the bottleneck speed,
because total amount of sent data among all bottlenecked flows is always acknowledged at bottleneck rate.
Besides, the traffic arriving rate is directly reflected by the slope of queue length increasing.
Under different traffic scenarios, dynamics of slope can be summarized as following laws:
\par \emph{Law 1} Without background flows, the slope in Phase 2 is equal to the port speed.
\par \emph{Law 2} When there is one background flow, or several background flows are congested at the same hop,
the slope in Phase 2 is smaller than the port speed.
\par \emph{Law 3} When several background flows are congested at previous hop,
slope in Phase 2 might be larger than port speed, but no larger than $2aR/(a+R)$,
where $a$ is the arriving rate of 1st round packets and $R$ is the port speed.
\par Moreover, from a microscopic perspective,
packet arrival pattern in forward path is evenly scheduled by ACK packets in reverse path.
Specifically, data packets from different sources are sent at full rate to bottleneck link.
After they arrive at the receiver, the ACKs will evenly spread in the reverse path of bottleneck link.
After these ACKs are fanned out to multiple senders,
they will evenly trigger the transmissions of data packets into network.
As a result, in the bottleneck queue, the queue length increases smoothly.
\par In conclusion, our experimental observations reveal that the bottleneck link and the self-clocking system are playing an important role in the evolution of micro-burst traffic.
\subsection{Implications}
\par These experimental observations have some implications:
\par \textbf{(1) Slope is an important indicator to mitigate micro-burst traffic.}
Slope directly reflects the bursty degree of aggregated traffic.
Specifically, the slope in Phase 1 reflects the flow concurrency,
and the slope in Phase 2 reflects the mismatch between congestion window increasing rate and bottleneck rate.
By carefully slowing down senders according to the slope, the sharp queue length increasing caused by micro-burst traffic can be effectively suppressed.
\begin{figure}[!t]
	\centering
	\subfigure[Queue length evolution]{
		\includegraphics[width=0.46\linewidth]{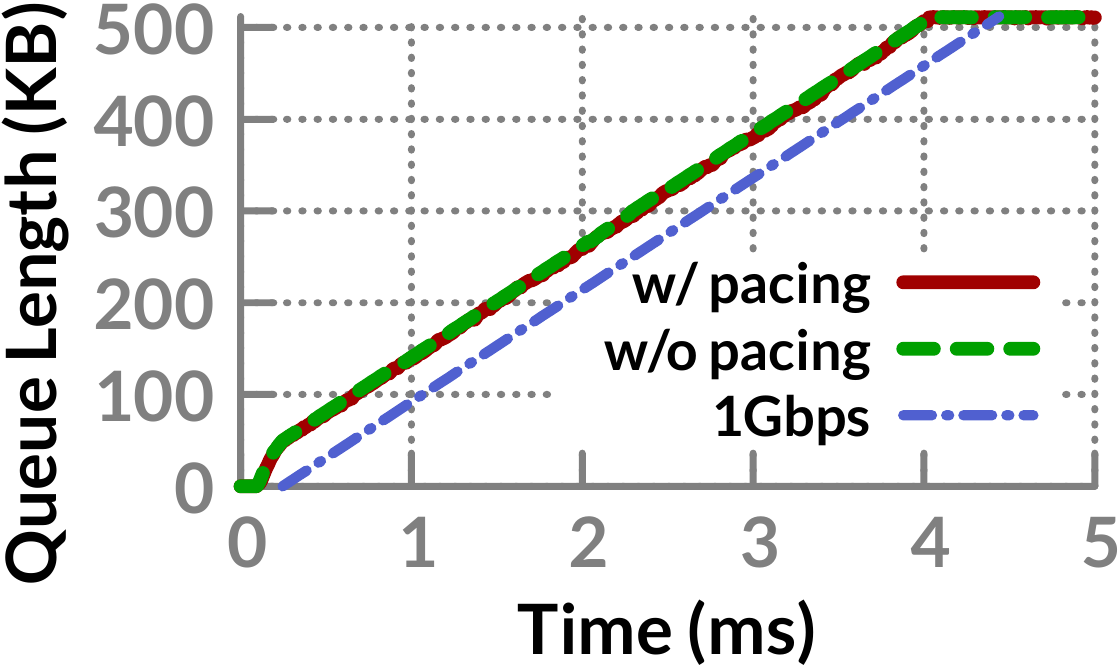}
		\label{fig:qevo-pacing}
	}
	\hfil
	\subfigure[Incast performance]{
		\includegraphics[width=0.46\linewidth]{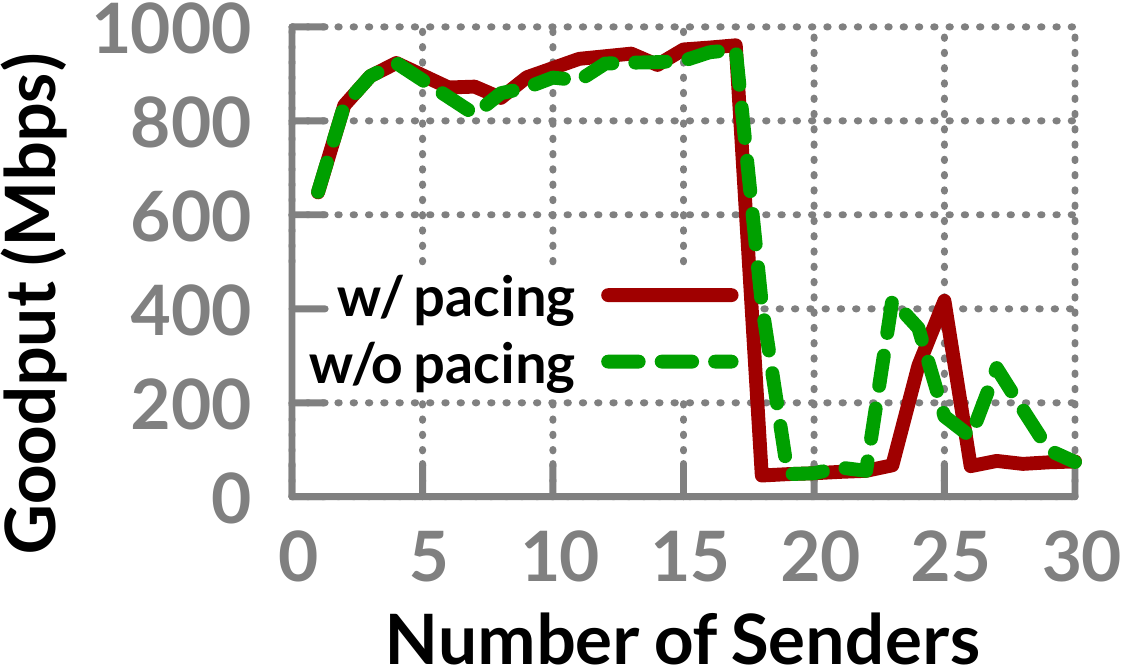}
		\label{fig:incast-pacing}
	}
	\caption{ Performance of TCP pacing.
		In the simulation, 40 servers are connected through the same switch.
		Experiment workload:
		(a) The same as that in \textsection \ref{section:characterizing:syn}.
		(b) Each sender sends 64KB data to the same receiver.
		Switch buffer size is 128KB.
	}
	\label{fig:ns2-pacing}
\end{figure}
\par \textbf{(2) TCP pacing does not help ease the sharp queue length increasing caused by fan-in traffic.}
In the literature \cite{SIGCOMM91ACK-compression, INFOCOM00Pacing, TCP-pacing-revisited, SIGMETRICS05JIANG},
TCP pacing is considered as an effective method to smooth traffic burstiness in a single flow
by evenly spreading a window of packets over a round-trip time.
However, in data centers with fan-in traffic, traffic burstiness comes from multiple flows.
Besides, transmissions of data packets have already been evenly triggered by ACK packets
(as is analyzed in \textsection \ref{section:characterizing:syn}).
Thus TCP pacing has little effect on queue length evolution,
and may not alleviate the impairments caused by micro-burst traffic (such as TCP incast throughput collapse). 
To validate this, we conduct simulations on ns-2 platform \cite{ns-2, ns2-pacing}.
The results are shown in \figurename{\ref{fig:ns2-pacing}}.
\par \textbf{(3) Simply absorbing micro-burst traffic may not help avoiding packet dropping,
	but partly magnifies micro-burst traffic.}
Senders won't slow down before they receive congestion signals.
If packets are absorbed without notifying senders of congestion,
the senders will believe that network is still fine and introduce more packets into network.
Specifically, if the sender is in slow start phase, absorbing every packet will introduce another two packets into network.

\subsection{Discussion} \label{section:discussion}
\noindent \textbf{Impact of other micro-burst traffic.}
\begin{figure}[!t]
	\centering
	\includegraphics[width=0.8\linewidth]{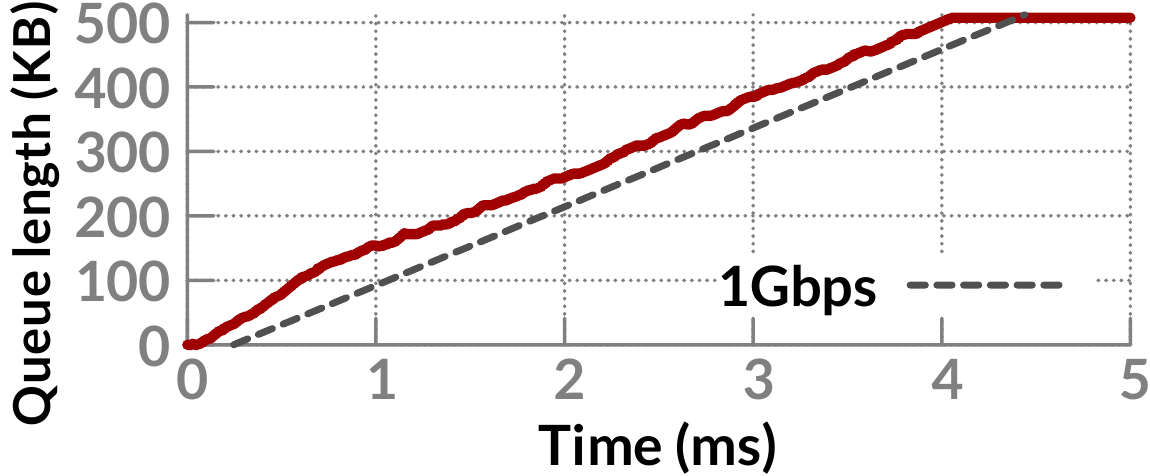}
	\caption{Queue length evolution when Large Segment Offload is turned on.}
	\label{fig:qevo-many-to-one-tso}
\end{figure}
One might ask whether other traffic burstiness (e.g., caused by LSO) affects our observations.
We find that these kinds of micro-burst traffic can only cause queue jitter at very small timescales,
because they disrupt the even transmission of data packets.
For example, we turn on LSO,
and use the same experiment settings as  \textsection \ref{section:characterizing:syn},
As shown in \figurename{\ref{fig:qevo-many-to-one-tso}},
the queue length evolution is similar to that without LSO.
\\ \textbf{Is these observations specific to TCP?}
Our most phenomena can also be observed in other TCP-like protocols,
because they use the same self-clocking mechanism
and similar congestion window adjustment algorithm as TCP.
Besides, the hidden buffer is caused by persistent buffer occupancy from long flows.
Since most protocols can not promise zero buffer occupancy,
hidden buffer also exists in other protocols.
However, current data center protocols, such as DCTCP \cite{SIGCOMM10DCTCP} and ECN* \cite{CoNEXT12ECN}, usually keep buffer occupancy low,
the effect of hidden buffer might be not as serious as observed in our experiments.

\section{Mitigating Micro-burst} \label{section:solution}
From previous experiments, we find that to mitigate micro-burst traffic,
senders need to slow down in time.
In this section,
we discuss the existing solutions that can potentially mitigate micro-burst traffic,
and show their limitations.
After that, we propose S-ECN scheme,
which can effectively suppress sharp queue length increasing caused by micro-burst traffic,
and reduce hidden buffer by keeping low buffer occupancy caused by long-lived flows.
\subsection{ECN Marking} \label{section:solution:exist-solution}
ECN \cite{rfc3168} can explicitly notify senders of congestion before buffer overflows.
Therefore, by using ECN, packet dropping caused by micro-burst traffic may be avoided.
Currently, ECN is widely used by Active Queue Management (AQM)  schemes (such as RED \cite{TON93RED} and PI \cite{INFOCOM01PI})
and some specific protocols (such as DCTCP \cite{SIGCOMM10DCTCP} and ECN* \cite{CoNEXT12ECN}).
In these schemes, packets are marked according to queue length.
For example, RED marks packets according to the average queue length.
DCTCP and ECN* mark packets according to instant queue length.
\par However, queue-length-based ECN marking has limitations when micro-burst traffic causes sharp queue length increasing,
because packets are not marked until the queue length reaches a threshold.
After queue length reaches the threshold, it takes time for all senders to receive the feedbacks from switch queue.
During this time, the queue length still keeps sharply increasing
and buffer may overflow before all senders slow down.
For example, assume that packets are marked with ECN when the instant queue length is larger than the threshold.
Multiple flows start concurrently and they share the same bottleneck.
At the beginning, all flows are in slow start phase,
therefore, the queue length will be increasing at port's sending rate (denoted by $R$).
When queue length reaches the ECN threshold (denoted by $h$),
the 1st sender will decrease its congestion window after $h/R + RTT$.
During this period, the queue length will increase to $h + R \cdot (h/R + RTT) = 2h + R \cdot RTT$.
Note at this time only the first sender slows down,
and it takes much more time for all senders to slow down.
We conduct an experiment when there are 9 senders
(more details in \textsection \ref{section:solution:evaluation}).
\figurename{\ref{fig:solution:qevo-dctcp}} shows that the queue length does not stop increasing until it reaches 104KB with DCTCP protocol,
while ECN threshold is only 32KB.
On the other hand, the ECN threshold should not be too low, otherwise network cannot be fully utilized \cite{CoNEXT12ECN}.
Therefore, we need another scheme to rapidly suppress the sharp queue length increasing.
\subsection{Design of S-ECN}
To effectively suppress the micro-burst traffic, we propose S-ECN.
S-ECN is inspired by the implication of slope:
slope reflects the mismatch between the congestion window increasing rate and bottleneck rate,
thus sharp queue increasing can be suppressed by properly marking packets according to slope.
Specifically, we mark packets immediately when the queue length begins to increase.
Furthermore, the fraction of marked packets indicates how large the slope is:
the bigger the slope, the larger the fraction.
\par Specifically, when slope (denoted by $s$)  is equal to or even larger than the port speed (denoted by $R$, which is constant in general),
then flows are in slow start phase.
All packets are marked with ECN so that flows will slow down immediately.
When slope is lower than the port's speed,
then packets are marked at a probability of $s/R$.
Finally, when slope is lower than or equal to 0 (i.e., the queue length is not increasing),
none of packets are marked.
The marking probability (denoted by $Prob$)  as a function of slope $s$ is as equation (\ref{eq:mark-prob}).
\begin{equation}
	Prob = \left\{\begin{array}{cl}
		0, & s \leqslant 0, \\
		\frac{s}{R}, & 0 < s < R, \\
		1, & s \geqslant R
	\end{array}
	\right.
	\label{eq:mark-prob}
\end{equation}
\subsection{Implementation of S-ECN}
\begin{figure}[!t]
	\centering
	\includegraphics[width=\linewidth]{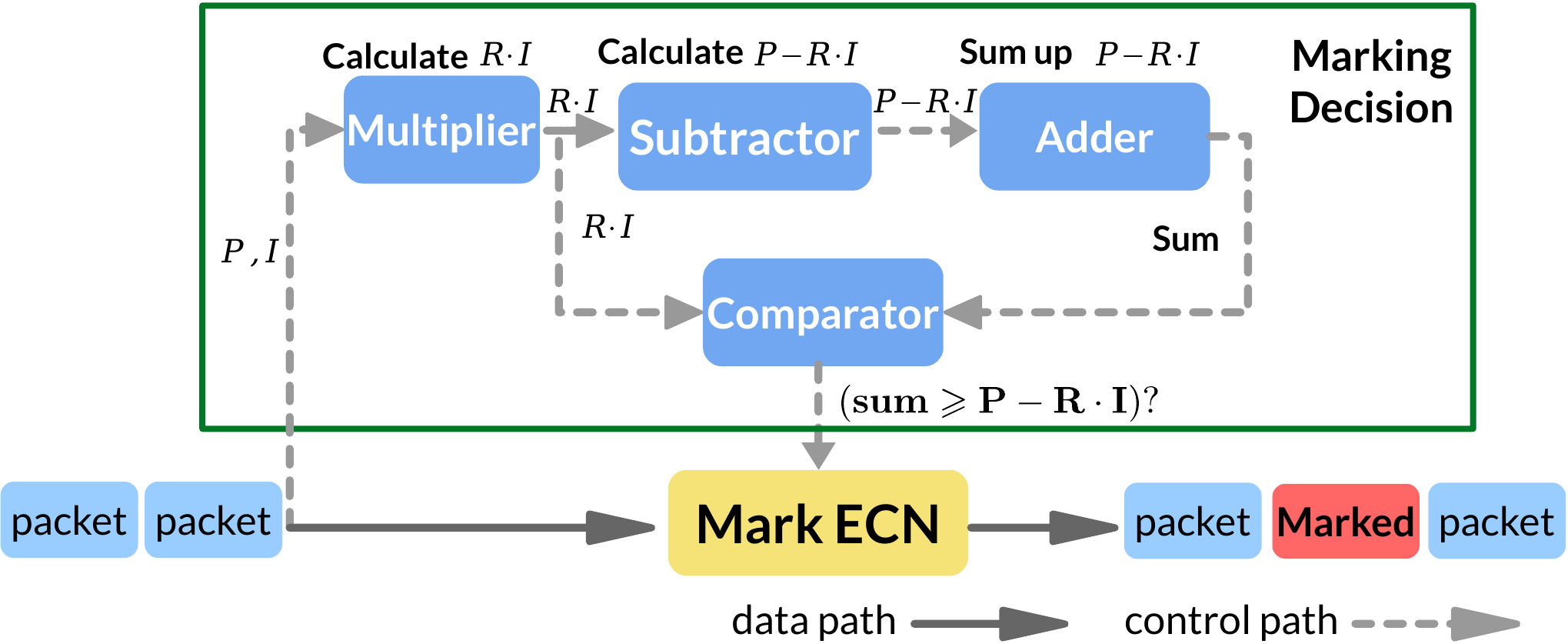}
	\caption{Implementation structure of S-ECN}
	\label{fig:secn:struct}
\end{figure}
We implement our scheme on NetFPGA platform \cite{NetFPGA}.
For simplicity, per-packet processing is used.
Specifically, in NetFPGA, we can easily get following information:
(i). Total size of an arriving packet (denoted by $P$, in bytes).
(ii). Time interval between arrivals of previous packet and the current packet (denoted by $I$, in clocks, 1 clock = 8ns).
(iii). Since our port rate (denoted by $R$, in Gbps) is constant, we can also directly use it.
Based on these information, we can calculate the marking probability given by equation (\ref{eq:mark-prob}).
Specifically, when a packet arrives at the switch's output port,
the current traffic arriving rate to the output port can be given by $P/I$.
Then the current slope is denoted by $s = P/I - R$.
If $0 < s < R$, the marking probability is $Prob = s/R = \frac{P - R \cdot I}{R \cdot I}$.
Finally, equation (\ref{eq:mark-prob}) can be rewritten as
\begin{equation}
	Prob = \left\{\begin{array}{cl}
		0, & P \leqslant R \cdot I, \\
		\frac{P - R \cdot I}{R \cdot I}, & R \cdot I < P < 2R \cdot I, \\
		1, & P \geqslant 2R \cdot I
	\end{array}\right.
\end{equation}
\par One method to implement marking with probability is by random number generator.
However, using random number generator will introduce large overhead in hardware.
Therefore we use another method:
instead of marking with probability $Prob$,
we can mark every $1/Prob$ packets.
Specifically, we can set up a threshold $T = \frac{R \cdot I}{P - R \cdot I}$
and there is a counter counting the number of unmarked packets.
When the number of unmarked packets is equal or larger than $T$,
then the next packet will be marked with ECN.
\par However, calculating the threshold needs to use divider, which will also introduce overhead.
In our implementation, we replace divider with adder and comparator.
Specifically, let $K$ be the number of unmarked packets, then 
an arriving packet is marked if $K \geqslant \frac{RI}{P - RI}$.
The condition can be rewritten as $K \cdot (P - RI) \geqslant RI$.
Instead of counting unmarked packets, we can add up $(P - RI$) for every arriving packet,
and compare the sum with $RI$.
In other words, the arriving packet is marked with ECN if $\sum_{j=1}^{K} (P_j - RI_j) > RI_K$.
\par Eventually, our implementation structure is shown in \figurename{\ref{fig:secn:struct}}.
When a packet arrives, we calculate $R \cdot I$ and $P - R \cdot I$.
Then $\sum_{j=1}^{K} (P_j - RI_j)$ is calculated by add $P - R \cdot I$	to accumulated sum
(i.e., $\textrm{sum} = \textrm{sum} + P - RI$).
Finally, the ECN marking decision is made by comparing $R \cdot I$ with the sum.
\tablename{\ref{tab:netfpga}} shows that our implementation needs less than 4\% extra resources in NetFPGA.
\subsection{Evaluation of S-ECN} \label{section:solution:evaluation}
In this section, we use realistic experiments in the testbed described in \textsection \ref{section:testbed} to show the performance of the S-ECN scheme.
\subsubsection{Protocols compared}
\begin{table}[!t]
	\centering
	\begin{tabular}{|l|c|c|c|} \hline
		{\bf Protocol} 	& {\bf End Host} & \multicolumn{2}{c|}{\bf Switch} \\ \cline{3-4}
		{\bf Name} 		&{\bf Algorithm} & {\bf S-ECN}	& {\bf ECN} \\ \hline \hline
		TCP				& TCP			& OFF 			& OFF \\
		ECN*			& TCP			& OFF			& ON \\
		S-ECN			& TCP			& ON			& OFF \\
		SL-ECN			& TCP			& ON			& ON \\ \hline
		DCTCP			& DCTCP			& OFF			& ON \\
		DCTCP+SL-ECN	& DCTCP			& ON			& ON \\ \hline
	\end{tabular}
	\caption{Protocols used in evaluation}
	\label{tab:solution:proto}
\end{table}
\par S-ECN scheme can work with multiple existing end host congestion control algorithms.
Specifically, we consider 6 protocols
(these protocols are listed in \tablename{\ref{tab:solution:proto}} for the sake of brevity):
\\ {\bf (i) TCP:}
End hosts use TCP NewReno as their congestion control algorithms.
Delayed ACK is disabled \cite{NSDI11YU}.
Switch uses taildrop queue and will not mark any packets with ECN.
\\ {\bf (ii) ECN*:}
The protocol proposed in \cite{CoNEXT12ECN}.
End hosts use TCP NewReno as their congestion control algorithms,
with ECN support (\texttt{/proc/sys/net/ipv4/tcp\_ecn = 1}).
The switch settings are similar as \cite{CoNEXT12ECN}.
Specifically, a packet will be marked with ECN
when it enters into the output queue and the queue length is larger than ECN threshold.
ECN threshold is set to 32KB.
\\ {\bf (iii) S-ECN:}
End hosts settings are the same as those of ECN.
Switch use S-ECN scheme to mark packets.
\\ {\bf (iv) SL-ECN:}
S-ECN can suppress the sharp queue length increasing, but may not keep queue length low.
Therefore, we introduce SL-ECN protocol, which combines queue-length-based and slope-based ECN marking schemes.
Specifically, end host settings are the same as those of ECN*.
In the switch, when the queue length is lower than the ECN threshold,
switch uses S-ECN scheme to mark packets.
When the queue length is higher than the ECN threshold,
all packets are marked with ECN.
ECN threshold is set to 32KB.
\\ {\bf (v) DCTCP:} The protocol proposed in \cite{SIGCOMM10DCTCP}.
We set $g = 0.125$.
Other settings are the same as those of ECN*,
including settings of switch.
\\ {\bf (vi) DCTCP+SL-ECN:}
End host settings are the same as those of DCTCP.
Switch settings are the same as those of SL-ECN.
\subsubsection{Microbenchmarks}
\begin{figure}[!t]
	\centering
	\subfigure[TCP]{
		\includegraphics[width=0.46\linewidth]{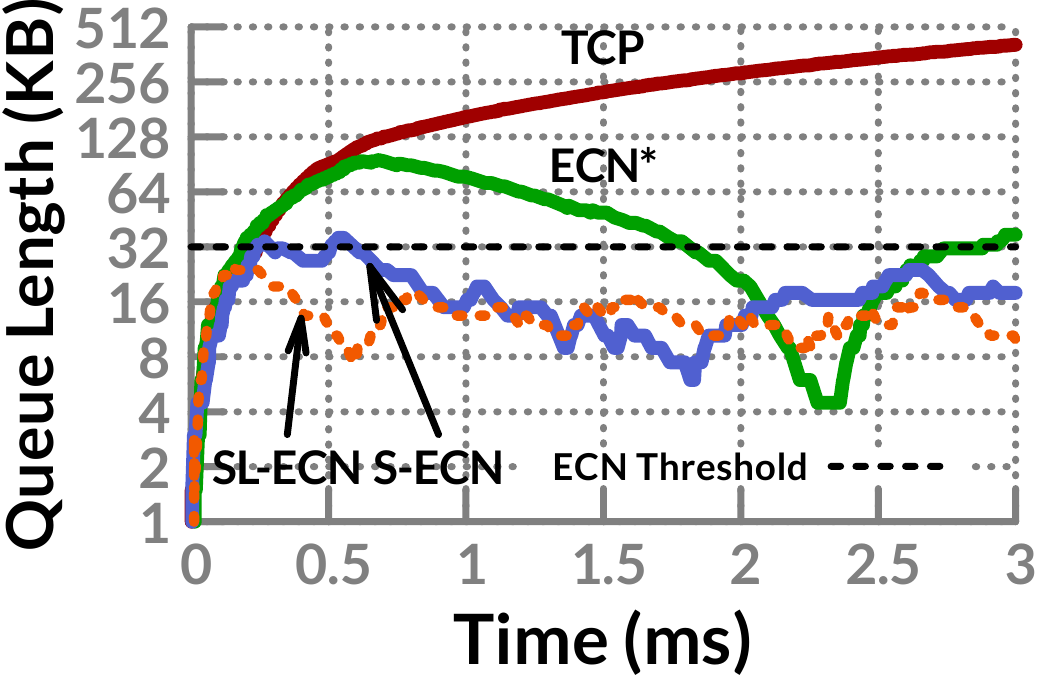}
		\label{fig:solution:qevo-tcp}
	}
	\hspace{-0.1in}
	\subfigure[DCTCP]{
		\includegraphics[width=0.48\linewidth]{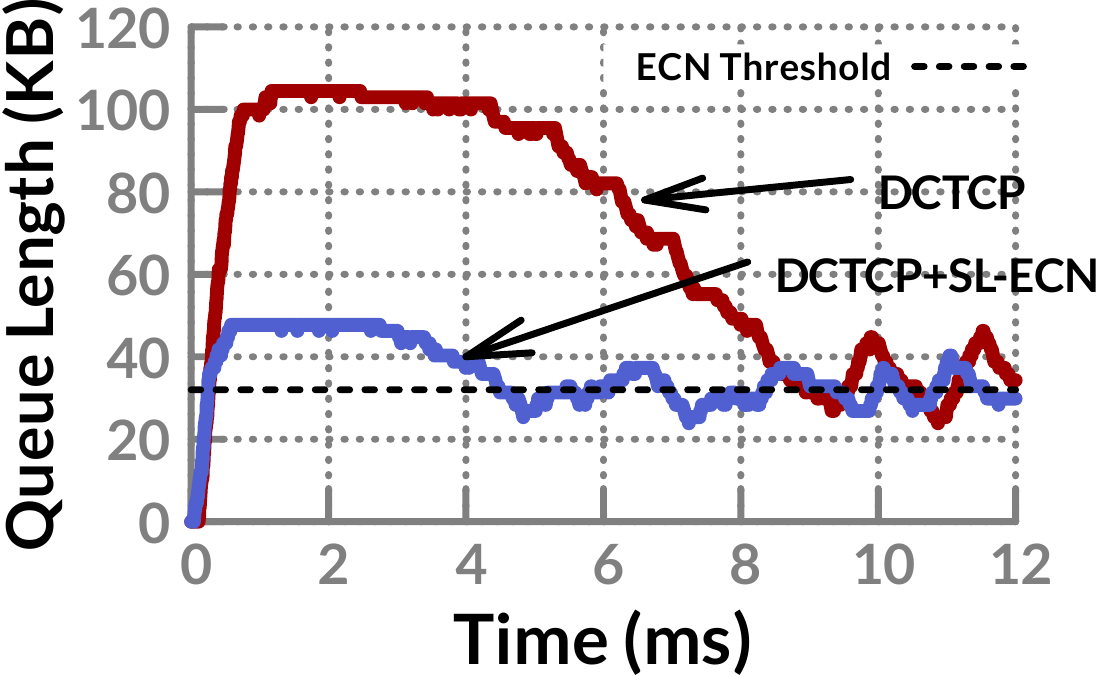}
		\label{fig:solution:qevo-dctcp}
	}
	\caption{Queue length evolution with difference protocols}
	\label{fig:solution:qevo}
\end{figure}
\noindent {\bf Sharp queue increasing suppression:}
First, we show how our scheme can help suppress the sharp queue length increasing.
In the experiment, we set the buffer size to 512KB.
We let a master host (Host 10) query slave hosts (Host 1-9) for 20MB data and get the instant queue length in the congested port.
\par The queue length evolution of each protocol is shown in \figurename{\ref{fig:solution:qevo}}.
Protocols are divided into two categories
by whether end hosts use TCP (\figurename{\ref{fig:solution:qevo-tcp}}) or DCTCP (\figurename{\ref{fig:solution:qevo-dctcp}}) algorithm.
First, when end host uses TCP algorithm, there are 4 protocols according to the ECN settings of switch.
With TCP, the queue length will be sharply increasing until buffer overflows.
With ECN* protocol, the queue length will be sharply increasing even after the queue length exceeds the ECN threshold.
The maximum queue length can reach 95.5KB, which is almost three times as high as the ECN threshold (32KB).
The reason why the queue length can not be limited under ECN threshold has been analyzed in \textsection \ref{section:solution:exist-solution}.
With S-ECN protocol, packets are marked ECN as soon as the queue length begins to increase at the beginning.
Therefore the queue length can be quickly controlled.
The maximum queue length is only 35.8KB, reduced by 62.5\% compared with ECN* protocol.
Queue length evolution of SL-ECN protocol is similar as that of SL-ECN protocol,
since the queue length for S-ECN protocol is always lower than the ECN threshold,
Furthermore, when all flows are in congestion avoidance phase (i.e., after $t = 2$ms),
with S-ECN and SL-ECN, the queue length is more steadier than that with ECN*.
The standard deviation of queue length is 4.657KB and 3.383KB with S-ECN and SL-ECN, respectively.
While the standard deviation is 13.234KB with ECN protocol.
Besides, with S-ECN and SL-ECN, the queue length can keep low.
Therefore, the effect of hidden buffer on micro-burst traffic can also be mitigated.
When the end hosts use DCTCP algorithm,
there are 2 protocols (\figurename{\ref{fig:solution:qevo-dctcp}}).
With DCTCP, the queue length can sharply increase to 104KB\footnotemark~at the beginning,
although ECN threshold is only 32KB.
\footnotetext{The maximum queue length is higher than that of ECN* protocol,
	since DCTCP considers history when judging congestion level
	and its congestion window decreasing method is smoother.}
With DCTCP+SL-ECN the queue length can be controlled under 47KB,
reduced by 53.8\% compared with DCTCP.
Furthermore, when all flows are in congestion avoidance phase (i.e., after $t = 10$ms),
with DCTCP+SL-ECN, the queue length is more steadier than that with DCTCP.
The standard deviation of queue length is 3.298KB.
Compare with it, with DCTCP, the standard deviation is 6.798KB,
which is almost twice as large as that with DCTCP+SL-ECN.
Besides, as DCTCP+SL-ECN keeps queue length low,
the effect of hidden buffer on micro-burst traffic can also be mitigated.
\begin{figure}[!t]
	\centering
	\includegraphics[width=3in]{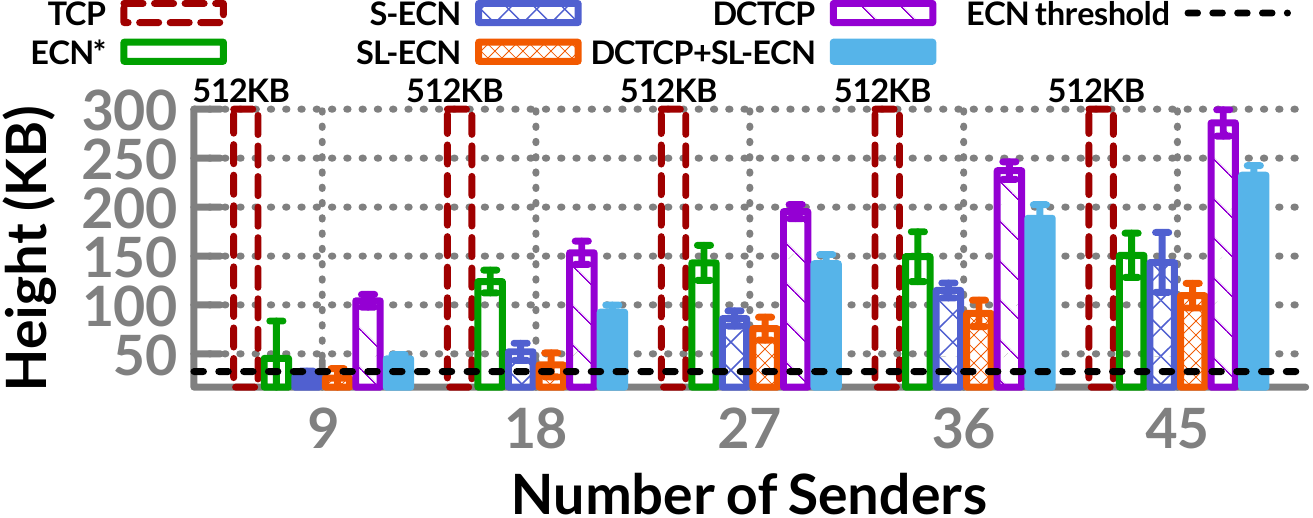}
	\caption{Maximum height the queue length in the congested port can reach.
		Maximum height of TCP is always 512KB because queue length will not stop increasing until buffer overflows.}
	\label{fig:solution:max-height}
\end{figure}
\par To show the maximum height the queue length can reach when there are more slaves,
we enlarge the number of slaves by letting each sending host (i.e., Host 1-9) emulate multiple slaves.
We re-conduct the previous experiment when the number of slaves is 9, 18, 27, 36, and 45,
and each experiment is repeated for 100 times.
\figurename{\ref{fig:solution:max-height}} shows the maximum height of each protocol.
Protocols using S-ECN scheme in the switch
can well control the queue length when the number of slaves is large.
For example, when there are 36 slaves, compared with ECN*, the maximum queue length can be reduced by 23.5\% (S-ECN) and 38.9\%(SL-ECN).
Compared with DCTCP, the maximum queue length can be reduced by 20.7\% with DCTCP+SL-ECN.
\begin{figure}[!t]
	\centering
	\includegraphics[width=3in]{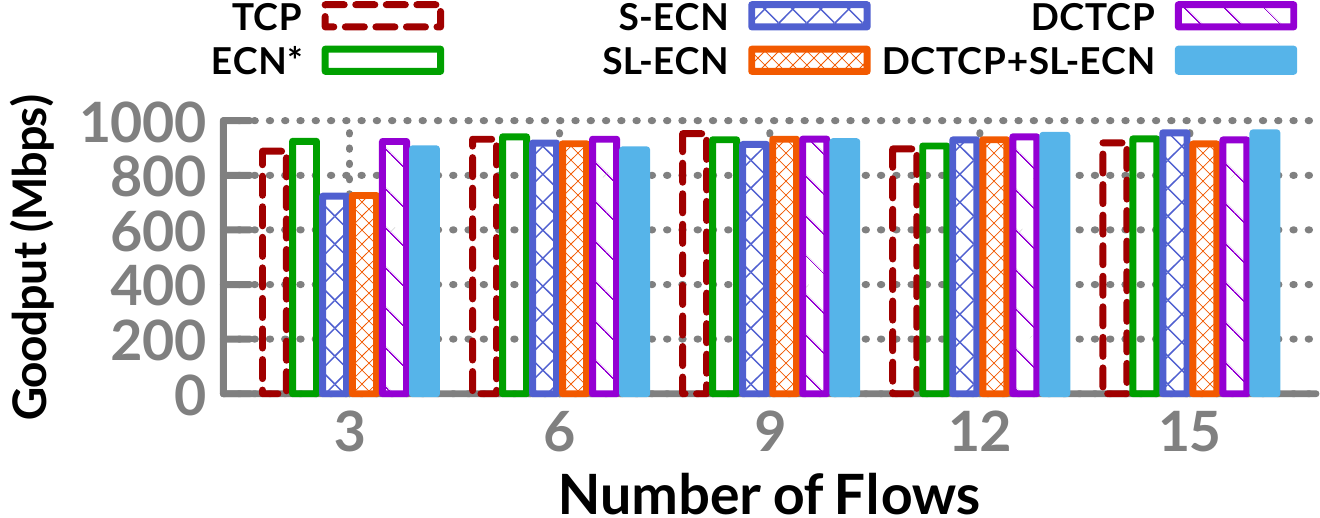}
	\caption{Goodput when the number of flows is small}
	\label{fig:solution:uti-goodput}
\end{figure}
\\ \noindent {\bf Network utilization:}
In this part, we evaluate the protocols when there are small number of long-lived flows.
To show this, we start 3 flows every 1 second,
who are from different senders and destine for the same receiver.
Each flow will send data as fast as it can after it starts.
\par We obtain the total throughput among all flows when a new bunch of flows start,
which is shown in \figurename{\ref{fig:solution:uti-goodput}}.
Throughput of DCTCP+SL-ECN is always over 900Mbps.
Thus, link utilization is nearly 100\%.
However, Throughput of S-ECN and SL-ECN is only 723Mbps and 726Mbps, respectively, when there are 3 flows.
This is because with these protocols sender will half its congestion window whenever it receives an ECN feedback,
which is too much.
Since with S-ECN scheme the fraction of marked packets is proportional to slope,
it can work better when senders cut its congestion window according to the fraction of marked packets (e.g., DCTCP).
\begin{figure}[!t]
	\centering
	\subfigure[TCP]{
		\includegraphics[width=0.46\linewidth]{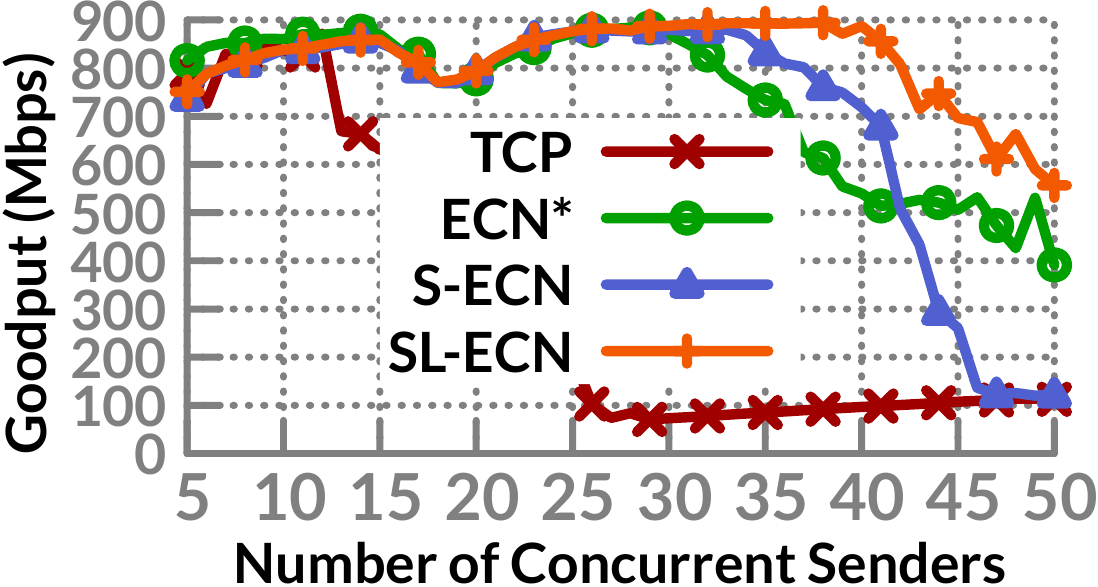}
		\label{fig:solution:goodput-tcp}
	}
	\hspace{-0.1in}
	\subfigure[DCTCP]{
		\includegraphics[width=0.46\linewidth]{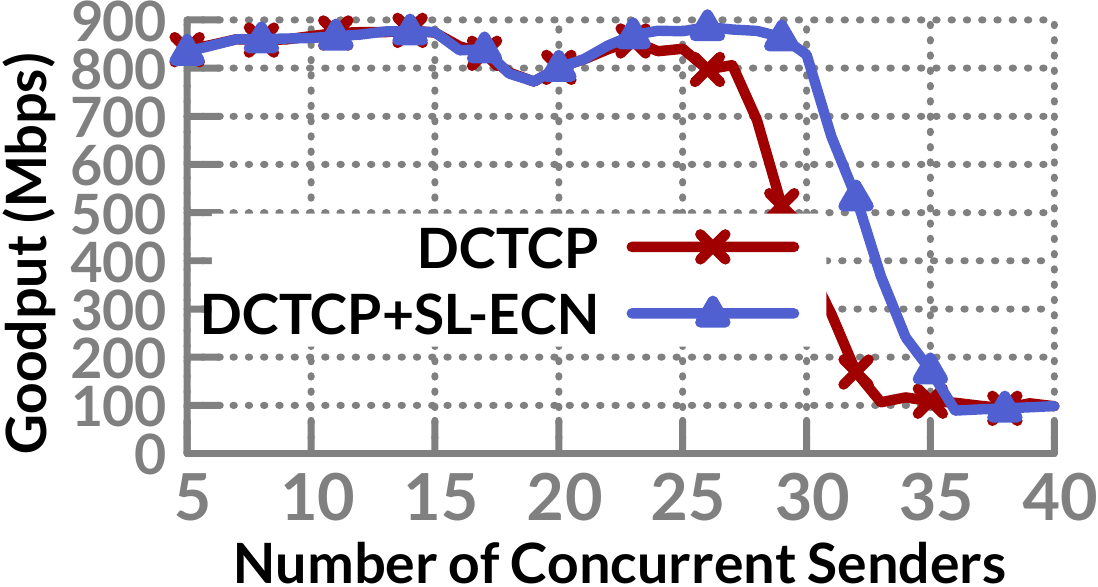}
		\label{fig:solution:goodput-dctcp}
	}
	\caption{Goodput when each slave respond with 64KB data}
	\label{fig:solution:incast-mapreduce}
\end{figure}
\begin{figure}[!t]
	\centering
	\subfigure[TCP]{
		\includegraphics[width=0.46\linewidth]{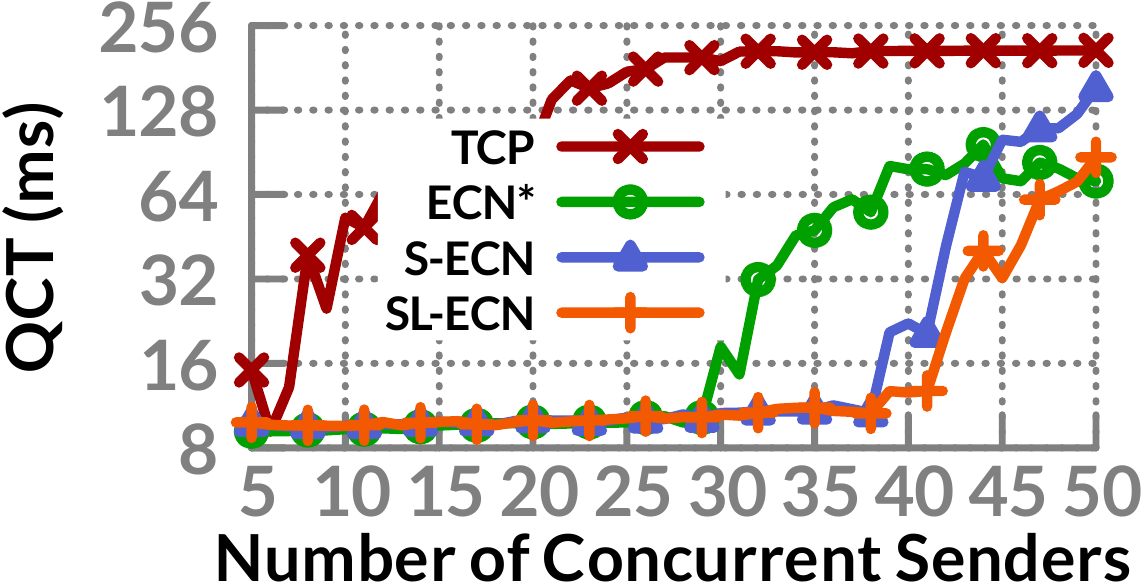}
		\label{fig:solution:qct-tcp}
	}
	\hspace{-0.1in}
	\subfigure[DCTCP]{
		\includegraphics[width=0.46\linewidth]{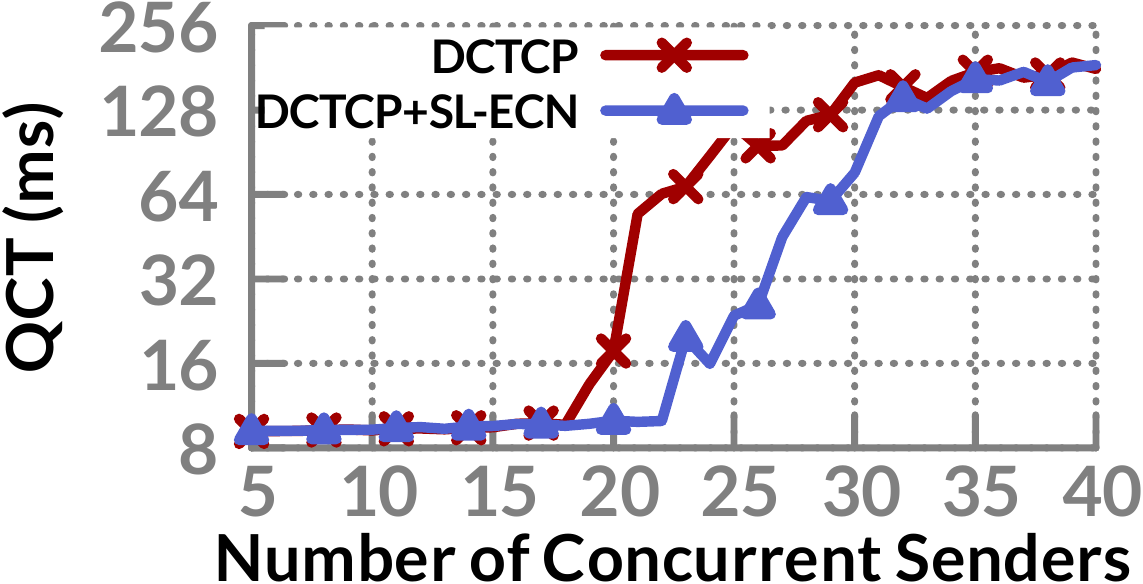}
		\label{fig:solution:qct-dctcp}
	}
	\caption{Query Completion Time (QCT) when each slave respond with $1024/n$ KB data.
		Note that y axis is log-scaled.}
	\label{fig:solution:incast-search}
\end{figure}
\\ {\bf Incast performance:}
By using S-ECN scheme, flows can avoid massive packet dropping caused by rapid queue length increasing,
and thus achieve high throughput.
In this part, we'll show how S-ECN scheme improves incast performance.
\par We consider two cases.
The 1st case is common in MapReduce applications.
A master server (Host 10) will send a query to other slaves (Host 1-9).
Each slave will respond with 64KB data.
Each host is used to emulate multiple senders \cite{SIGCOMM11D3}.
The buffer size is 128KB.
\figurename{\ref{fig:solution:incast-mapreduce} shows the goodput when the number of slaves varies from 2 to 50.
	In \figurename{\ref{fig:solution:goodput-tcp}}, end hosts use TCP algorithm.
	S-ECN outperforms ECN* when there are less than 40 hosts.
	However, when the number of hosts is larger than 40, S-ECN is worse than ECN*.
	This is because S-ECN can control the slope (i.e., queue length increasing rate),
	but can not decrease the queue length when there are large number of senders.
	Compared with it, SL-ECN protocol considers both slope and queue length.
	It can support over 40 senders.
	While ECN* protocol can support about 30 senders.
	In \figurename{\ref{fig:solution:goodput-dctcp}}, end hosts use DCTCP algorithm.
	All protocols can achieve the same goodput when the number of senders are small.
	Before TCP incast happens, DCTCP protocol can support ~25 senders in our experiment,
	while DCTCP+SL-ECN protocol can support as many as 30 senders.
	\par The 2nd case is common in web search applications
	The master (Host 10) will query other slaves (Host 1-9) for 1024KB data in all.
	In other words, each slave will respond with $1024/n$ KB data.
	\figurename{\ref{fig:solution:incast-search}} shows the query completion time for each protocol,
	where we have similar observations as previous experiment.
	\subsubsection{Benchmark traffic}
	\begin{figure}[!t]
		\centering
		\includegraphics[width=0.9\linewidth]{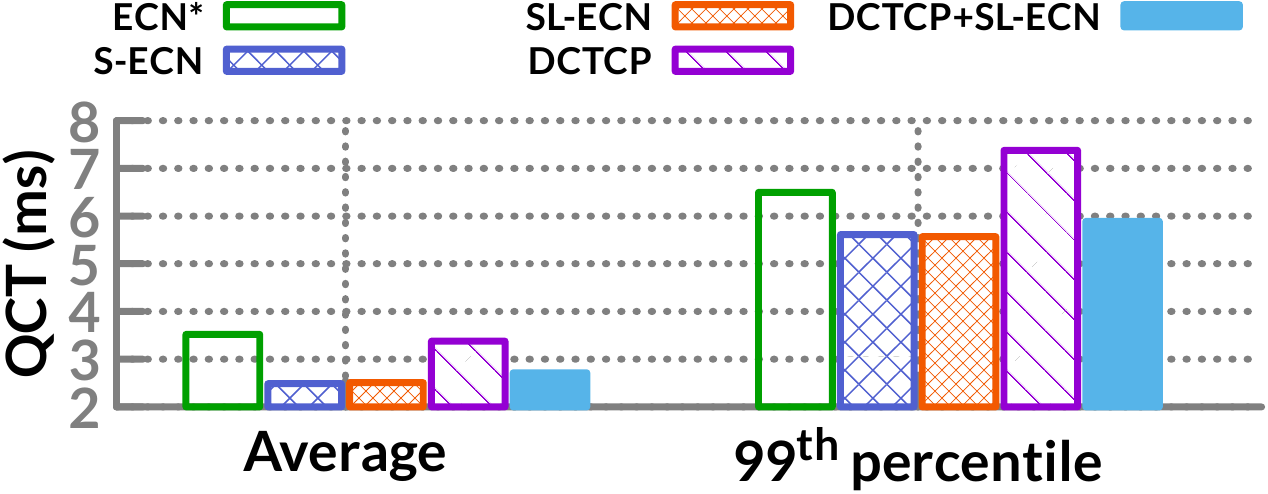}
		\caption{Query completion time.}
		\label{fig:solution:benchmark:qct}
	\end{figure}
	\begin{figure}[!t]
		\centering
		\subfigure[$(0, 100\textrm{KB}\rbracket$]{
			\includegraphics[height=0.9in]{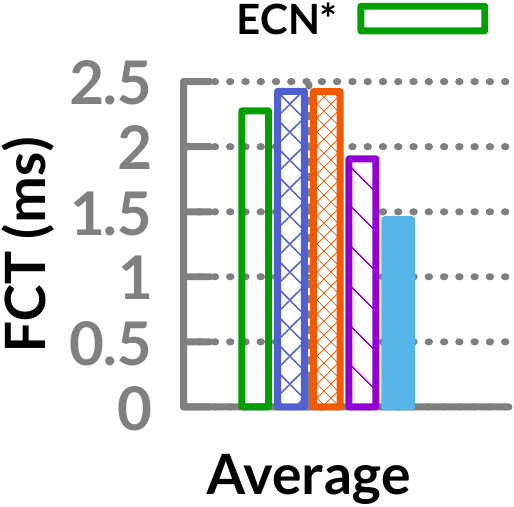}
		}
		\hfil
		\subfigure[$(0, 100\textrm{KB}\rbracket$]{
			\includegraphics[height=1in]{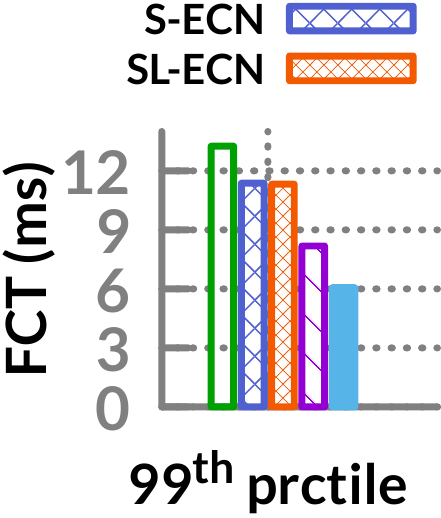}
		}
		\hfil
		\subfigure[$(10\textrm{MB}, +\infty)$]{
			\includegraphics[height=1in]{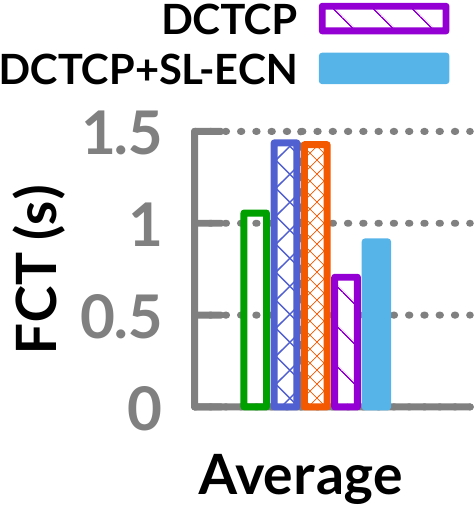}
			\label{fig:solution:benchmark:ws-fct-10MB}
		}
		\caption{Flow completion time across different flow sizes.}
		\label{fig:solution:benchmark:ws-fct}
	\end{figure}
	\begin{figure*}[!t]
		\centering
	\end{figure*}
	\noindent \textbf{Experiment settings:}
	We use a realistic workload to evaluate our scheme,
	which is derived from data center supporting web search service \cite{SIGCOMM10DCTCP}.
	In the workload, there are two kinds of traffic between servers.
	One is query traffic, where a server will periodically send a query to all other servers
	and each server will send a response message back after receiving the query message.
	The total response size is 100KB.
	Query arrivals follow a Poisson process.
	The other kind of traffic is background traffic, which follows a one-to-one pattern.
	The flow size distribution is from \cite{SIGCOMM10DCTCP}.
	The flow arrivals in each server follow a Poisson process.
	The ratio of query traffic and background traffic is drawn from \cite{SIGCOMM10DCTCP},
	and network load in our experiment is 0.4.
	In our experiments, switch buffer of each port is 128KB,
	and RTO$_{\textrm{min}}$ is set to 10ms.
	Each experiment lasts for 5 minutes.
	Over 350,000 query flows and 140,00 background flows are generated, respectively.
	\\ \textbf{Experiment results:}
	By suppressing the fast queue increasing, flows can finish faster.
	\figurename{\ref{fig:solution:benchmark:qct}} shows the query completion time in web search workload\footnotemark.
	Compared to ECN* protocol, S-ECN and SL-ECN can reduce the average query completion time by 29.4\% and 28.7\%, respectively,
	and they can reduce the $99^{th}$ query completion time by 13.6\% and 14.3\%, respectively.
	Compared to DCTCP protocol, DCTCP+SL-ECN can reduce the average query completion time by 19.6\%,
	and $99^{th}$ query completion time is reduced by 20.1\%.
	\par Small flows can benefit from suppressing the fast queue increment caused by query traffic.
	\footnotetext{Since the performance of TCP is outside the plot range, it is omitted in the figure.}
	\figurename{\ref{fig:solution:benchmark:ws-fct}} (a,b) shows the average and $99^{th}$ percentile completion time of flows whose size is smaller than 100KB.
	Compared to ECN* protocol, S-ECN and SL-ECN do not improve the average flow completion time,
	but they can reduce the $99^{th}$ percentile of flow completion time by 14.2\% and 14.5\%, respectively.
	Compare to DCTCP protocol, DCTCP+SL-ECN can reduce the average flow completion time by 24.5\%
	and reduce the $99^{th}$ percentile by 25.9\%.
	\par However, S-ECN scheme still needs to be improved.
	As shown in \figurename{\ref{fig:solution:benchmark:ws-fct-10MB}},
	S-ECN, SL-ECN, and DCTCP + SL-ECN protocol do not perform well for large flows.
	This is part of our future work.

\section{Related Work} \label{section:related-work}
\par Lots of studies investigate the burstiness in wide area networks.
In \cite{SIGCOMM91ACK-compression}, authors find that ACK packets can be bunched up when encountering queuing,
which in turn causes bursty transmission at sender.
The phenomenon is called ACK-compression.
The study \cite{ToN99Reorder} finds that ACK reordering can result in traffic burstiness.
\cite{INFOCOM00Pacing} outlines burstiness in TCP and investigates TCP pacing's performance.
\cite{IMC03Burst} identifies several causes of bursts from individual flows
and examine their effects.
\cite{SIGMETRICS05JIANG} examines how TCP creates burstiness in sub-RTT timescales,
and finds that TCP's self-clocking and network queuing can shape the packet interarrivals into two-level ON-OFF pattern.
\cite{PAM05Allman} assesses the presence and impact of TCP burstiness through analyzing real network traces,
and finds that large bursts can always cause packet loss but they rarely occur.
Several burst mitigation methods,
including MaxBurst, Aggressive Maxburst, User It or Lose It, Congestion Window and Slow Start Threshold Limiting,
are studied in \cite{CCR05Allman}. 
\cite{PAM05burst} studies the burstiness of TCP flows at packet level.
However, these studies mainly focus on burstiness in a single flow.
While we study the behavior of micro-burst traffic when many flows aggregate in data center networks.
\par In data centers, traces from ten data centers is examined in \cite{IMC10DC},
and authors find that traffic exhibits an ON/OFF pattern.
\cite{CoNEXT13NIC} studies new causes of bursts,
including offloading features in NIC, batching schemes, and bursty OS system calls and APIs.
Several tools \cite{NSDI11MB, SIGCOMM14TPP} are developed to detect micro-bursts.
In \cite{HotNets15Micro}, authors find that short-lived congestion caused by micro-burst brings challenges to load balancing systems.
Different from these studies, we focus on the nature of micro-burst traffic introduced by transport protocol.
When micro-burst traffic occurs,
lots of solutions are proposed to address the TCP incast problem
\cite{SIGCOMM09incast, CoNEXT10incast, ICNP13GIP, ICNP14PAC},
and reduce flow completion time \cite{SIGCOMM10DCTCP, SIGCOMM11D3, SIGCOMM12D2TCP, SIGCOMM12DeTail, NSDI14CP}.
However, these solutions do not touch the essence of the micro-burst traffic.

\section{Conclusion} \label{section:conclusion}
Micro-burst traffic in data centers may cause packet dropping,
which can bring about serious performance degradation (e.g., TCP incast problem).
However, the root cause and dynamic behavior of micro-burst are not comprehensively studied
under data center network's unique communication patterns and topologies.
In this paper, we use real experiments to examine the micro-burst traffic through observing queue length at find-grained timescale in typical traffic scenarios.
We find that self-clocking mechanism and bottleneck link jointly dominate the evolution of micro-burst.
We also find that slope of queue length increasing can describe the dynamic behavior of micro-burst under all scenarios.
These findings implicates that traditional solutions like absorbing and pacing to mitigate micro-burst traffic
are not effective in data centers.
To suppress the micro-burst traffic, the sending rate need to be throttled as soon as possible.
Enlightened by our findings and implications, we propose S-ECN policy, which takes advantage of slope of queue length increasing
and randomly marks packets with ECN.
The experiment results show that protocols using S-ECN policy can effectively mitigate micro-burst traffic.

	\bibliographystyle{abbrv}
	\bibliography{mybibfile}
	\label{pages}
\end{document}